\documentclass[12pt,preprint]{aastex}

\font\tenpt=cmr10

\def\Prob   {\ifmmode {{\rm Prob}} \else {Prob} \fi}
\def\arm    {\ifmmode {{\rm arm}} \else {arm} \fi}

\def\twopi  {\ifmmode {2\pi}   \else {$2\pi$}   \fi}

\def\kms    {\ifmmode {{\rm km s}^{-1}} \else {km s$^{-1}$} \fi}

\def\uas  {\ifmmode {\mu{\rm as}}\else{$\mu$as}\fi}
\def\deg  {\ifmmode {^\circ}\else {$^\circ$}\fi}
\def\porm {\ifmmode {\pm}\else {$\pm$}\fi}

\def\chisqpdf {\ifmmode {\chi^2_{\rm pdf}}\else {$\chi^2_{\rm pdf}$}\fi}
\def\chisq    {\ifmmode {\chi^2}\else {$\chi^2$}\fi}

\def\Msun {M$_\odot$}
\def\HI   {H~{\small I}}
\def\Hi   {H~{\scriptsize I}}
\def\HII  {H~{\small II}}

\def\firstQ  {1$^{st}$}

\def\etal {et al.~}
\def\eg   {e.g.,~}
\def\ie   {i.e.,~}

\def\d    {\ifmmode {{\rlap{.}}^\circ}\else {${\rlap{.}}^\circ$}\fi}
\def\s    {\ifmmode {{\rlap{.}}^s}\else {${\rlap{.}}^s$}\fi}
\def\as   {\ifmmode {{\rlap{.}}^{''}}\else {${\rlap{.}}^{''}$}\fi}

\newbox\grsign \setbox\grsign=\hbox{$>$} \newdimen\grdimen \grdimen=\ht\grsign
\newbox\laxbox \newbox\gaxbox
\setbox\gaxbox=\hbox{\raise.5ex\hbox{$>$}\llap
     {\lower.5ex\hbox{$\sim$}}}\ht1=\grdimen\dp1=0pt
\setbox\laxbox=\hbox{\raise.5ex\hbox{$<$}\llap
     {\lower.5ex\hbox{$\sim$}}}\ht2=\grdimen\dp2=0pt
\def\gax{\mathrel{\copy\gaxbox}}
\def\lax{\mathrel{\copy\laxbox}}

\def\pa    {\ifmmode {\psi} \else {$\psi$}\fi}
\def\rPpm  {\ifmmode {r_{\Ro,\To}} \else {$r_{Ro,\To}$}\fi}

\def\vlsr  {\ifmmode {v}\else {$v$}\fi}
\def\vhelio{\ifmmode {v_{Helio}}\else {$v_{Helio}$}\fi}
\def\delV  {\ifmmode {\Delta v}\else {$\Delta v$}\fi}
\def\sigV  {\ifmmode {\sigma_v}\else {$\sigma_v$}\fi}

\def\ura   {\ifmmode {\mu_\alpha}\else {$\mu_\alpha$}\fi}
\def\udec  {\ifmmode {\mu_\delta}\else {$\mu_\delta$}\fi}

\def\ul    {\ifmmode {\mu_l}\else {$\mu_l$}\fi}
\def\ub    {\ifmmode {\mu_b}\else {$\mu_b$}\fi}

\def\uml   {\ifmmode {v_{gr}}\else {$v_{gr}$}\fi}
\def\umb   {\ifmmode {v_b}\else {$v_b$}\fi}
\def\vsrad {\ifmmode {v_{rad}}\else {$v_{rad}$}\fi}

\def\upl   {\ifmmode {v^p_{gr}}\else {$v^p_{gr}$}\fi}
\def\upb   {\ifmmode {v^p_b}\else {$v^p_b$}\fi}
\def\vprad {\ifmmode {v^p_{rad}}\else {$v^p_{rad}$}\fi}

\def\Vo    {\ifmmode {V^{Std}_\odot}\else {$V^{Std}_\odot$}\fi}
\def\Uo    {\ifmmode {U^{Std}_\odot}\else {$U^{Std}_\odot$}\fi}
\def\Wo    {\ifmmode {W^{Std}_\odot}\else {$W^{Std}_\odot$}\fi}
\def\VH    {\ifmmode {V^H_\odot}\else {$V^H_\odot$}\fi}
\def\UH    {\ifmmode {U^H_\odot}\else {$U^H_\odot$}\fi}
\def\WH    {\ifmmode {W^H_\odot}\else {$W^H_\odot$}\fi}
\def\V     {\ifmmode {V_\odot}\else {$V_\odot$}\fi}
\def\U     {\ifmmode {U_\odot}\else {$U_\odot$}\fi}
\def\W     {\ifmmode {W_\odot}\else {$W_\odot$}\fi}

\def\Vs    {\ifmmode {V_s}\else {$V_s$}\fi}
\def\Us    {\ifmmode {U_s}\else {$U_s$}\fi}
\def\Ws    {\ifmmode {W_s}\else {$W_s$}\fi}

\def\Vsbar {\ifmmode {\overline{V_s}}\else {$\overline{V_s}$}\fi}
\def\Usbar {\ifmmode {\overline{U_s}}\else {$\overline{U_s}$}\fi}
\def\Wsbar {\ifmmode {\overline{W_s}}\else {$\overline{W_s}$}\fi}

\def\aone  {\ifmmode {a_1}\else {$a_1$}\fi}
\def\atwo  {\ifmmode {a_2}\else {$a_2$}\fi}
\def\athr  {\ifmmode {a_3}\else {$a_3$}\fi}

\def\pars  {\ifmmode{\pi_s}\else{$\pi_s$}\fi}

\def\Ts    {\ifmmode{\Theta_s}\else{$\Theta_s$}\fi}
\def\Tdot  {\ifmmode{d\Theta\over dR}\else{$d\Theta\over dR$}\fi}

\def\Rp    {\ifmmode{R_p}\else{$R_p$}\fi}

\def\To    {\ifmmode{\Theta_0}\else{$\Theta_0$}\fi}
\def\Ro    {\ifmmode{R_0}\else{$R_0$}\fi}
\def\Ho    {\ifmmode{H_0}\else{$H_0$}\fi}

\def\lbv     {\ifmmode {(l,b,v)}\else{$(l,b,v)$}\fi}
\def\lv      {\ifmmode {(l,v)}\else{$(l,v)$}\fi}
\def\lvS     {\ifmmode {(l,v)_{\rm src}}\else{$(l,v)_{\rm src}$}\fi}
\def\lbvS    {\ifmmode {(l,b,v)_{\rm src}}\else{$(l,b,v)_{\rm src}$}\fi}
\def\lbvA    {\ifmmode {(l,b,v)_{\rm arm}}\else{$(l,b,v)_{\rm arm}$}\fi}
\def\lbvRBD  {\ifmmode {(l,b,v,R,\beta,d)}\else{$(l,b,v,R,\beta,d)$}\fi}

\def\Nbins   {\ifmmode{N_{\rm bins}}\else{$N_{\rm bins}$}\fi}
\def\DelD    {\ifmmode{\Delta d}\else{$\Delta d$}\fi}

\slugcomment{2016 April 1}
\shorttitle{Distances to Sources in Spiral Arms} 
\shortauthors{Reid \etal}

\begin{document}

\title{A PARALLAX-BASED DISTANCE ESTIMATOR FOR SPIRAL ARM SOURCES
      }   

\author{M. J. Reid\altaffilmark{1}, T. M. Dame\altaffilmark{1}, 
        K. M. Menten\altaffilmark{2}, A. Brunthaler\altaffilmark{2} 
       }

\altaffiltext{1}{Harvard-Smithsonian Center for
   Astrophysics, 60 Garden Street, Cambridge, MA 02138, USA}
\altaffiltext{2}{Max-Planck-Institut f\"ur Radioastronomie, 
   Auf dem H\"ugel 69, 53121 Bonn, Germany}

\begin{abstract}
The spiral arms of the Milky Way are being accurately located for the first time 
via trigonometric parallaxes of massive star forming regions with the BeSSeL Survey, 
using the Very Long Baseline Array and the European VLBI Network, and 
with the Japanese VERA project. Here we describe a computer program that leverages 
these results to significantly improve the accuracy and reliability of distance 
estimates to other sources that are known to follow spiral structure. Using a 
Bayesian approach, sources are assigned to arms based on their \lbv\ coordinates 
with respect to arm signatures seen in CO and \HI\ surveys.  A source's kinematic 
distance, displacement from the plane, and proximity to individual parallax sources 
are also considered in generating a full distance probability density function. 
Using this program to estimate distances to large numbers of star forming regions, 
we generate a realistic visualization of the Milky Way's spiral structure as seen 
from the northern hemisphere.
\end{abstract}


\section{Introduction} \label{sect:introduction}

The Bar and Spiral Structure Legacy (BeSSeL) Survey
\footnote[3]{http://bessel.vlbi-astrometry.org}
 and the Japanese VLBI Exploration of Radio Astrometry (VERA) 
\footnote[4]{http://veraserver.mtk.nao.ac.jp}
are now supplying large numbers of trigonometric parallaxes to sites of
massive star formation across large portions of the Milky Way.
Since massive star forming regions (MSFRs) are excellent tracers of
spiral structure in galaxies, these parallax distances allow one to 
construct a sparsely sampled map of the spiral structure of the Milky Way
\citep{Honma:12,Reid:14}. This map can be used to characterize spiral arms, even if 
they are somewhat irregular \citep{Honig:15}, and to provide a predictive model for the 
locations of other sources that {\it a priori} are likely to reside in arms, 
such as \HII\ regions, giant molecular clouds (GMCs), and associated 
infrared sources.  

In this paper we describe a computer program and web-based application for
estimating distances to spiral arm sources based solely on their Galactic longitude 
and latitude coordinates and local standard-of-rest velocities: \lbvS.  
In many cases, one can use a source's \lbvS\ coordinates to confidently assign 
it to a portion of a spiral arm, based on its proximity to \lbvA\ traces of 
spiral arms seen in CO or \HI\ surveys.  Since distances to large sections of arms are 
now known \citep{Xu:13,Zhang:13,Choi:14,Sato:14,Wu:14,Hachisuka:15,Xu:16}, 
this can lead to an accurate and reliable distance estimate.  
We also add information from a kinematic model, Galactic latitude,
and proximity to individual giant molecular clouds for which parallax distances 
have been measured. 
Since spiral arms are not circular, but wind outward from the Galactic center 
with pitch angles between about 7\deg\ and 20\deg\ \citep{Reid:14}, 
arm location and kinematic distance information can be complementary.   
We combine all of this information, using a Bayesian approach similar to that
introduced by \citet{Ellsworth-Bowers:13} for resolving kinematic distance
ambiguities, to estimate a probability density function (PDF) for source distance.   

Details of the distance PDF calculation are given in Section 2 and some example 
PDFs are discussed in Section 3.  In Section 4, we compare our program's performance
on a sample of \HII\ regions thought to be very distant based on \HI\ absorption
information.  In Section 5, we estimate distances of large 
numbers of spiral arm sources from Galactic plane surveys to provide a visualization 
of the Milky Way, which includes realistic distributions of star formation activity.  
Finally, the Appendix displays the Galactic \lv\ plots, used to determine the \lbvA\ 
traces for spiral arms, which are central to determining distance PDFs.

\section{Bayesian Distance Estimation} \label{sect:distance}

        Measurements of parallaxes of masers in massive star forming
regions allow one to reliably trace the locations of spiral arms in the 
Milky Way.   Since spiral arms can be clearly identified as continuous
curves in \lbvA\ traces of \HI\ \citep{Weaver:70}
or CO \citep{Cohen:80} emission, one can often assign an individual source to a spiral arm 
based solely on its \lbvS\ coordinates.  Thus, combining the locations of spiral 
arms, from parallax measurements, with assignment to an arm, any source that is associated 
with spiral arms can be located in the Milky Way in three-dimensions with some degree 
of confidence. 

In addition to spiral arm assignment, there is other information
available that indicates distance, including kinematic distance, Galactic latitude,
and location within a giant molecular cloud with a measured  
parallax.  These types of distance information are subject to uncertainty 
in varying ways and can best be combined in a Bayesian approach.  
We employ this information by constructing a PDF for each type of 
distance information and then multiplying them together to arrive at a combined 
distance PDF.  With simplified notation, the probability density that a source
is at a distance $d$ is given by
$$\Prob(d) \propto \Prob_{SA}(d)\times\Prob_{KD}(d)\times\Prob_{GL}(d)\times\Prob_{PS}(d)~~,$$
where the subscripts indicate different types of distance information
($SA$: spiral arm model; $KD$: kinematic distance; 
$GL$: Galactic latitude; $PS$: parallax source) as detailed below.

We fit the combined distance PDF with a model of a flat background probability density 
and multiple Gaussian components, estimating their peak probability densities, locations 
(\ie distances), and widths (\ie distance uncertainties).
The total probability of each component is calculated by integrating its Gaussian 
probability density over distance.  Note that the component with the greatest 
integrated probability may not have the greatest probability {\it density}, since a tall, 
narrow Gaussian can have less area than a short, wide Gaussian.  The program reports
the component with the greatest integrated probability, and up to two more
components if significant. 

The BeSSeL Survey website ({\tt http://bessel.vlbi-astrometry.org}) provides a web-based 
application that generates distance PDF plots similar to those in Section \ref{sect:examples}.
We also make available the {\tenpt FORTRAN} program, which allows for more efficient 
use on large catalogs of star forming sources.  Also available are the detailed spiral arm 
\lbvRBD\ traces used by the program.  In addition to the Galactic longitude, latitude and
LSR velocities of the arm segments, they include current best estimates of the 
Galacticocentric radii ($R$), azimuths ($\beta$; defined as 0 toward the Sun and 
increasing clockwise as viewed from the North Galactic Pole), 
and heliocentric distances ($d$).

\subsection {Spiral Arm PDF}\label{sect:SA}

Key distance information is provided by the locations of spiral arm segments in the Galaxy, 
which are based largely on trigonometric parallaxes of high mass star forming regions.   
As mentioned above, we compare a target 
source's \lbvS\ with traces of an arm segment's \lbvA\ values to estimate the probability
that the source is associated with an arm segment.  In the Appendix, we provide
some background for the identification of spiral arms in \lv\ plots and
a description of the \lbv\ traces, which we determined from CO and \HI\ data.
 
\begin{figure}[ht]
\epsscale{0.65} 
\plotone{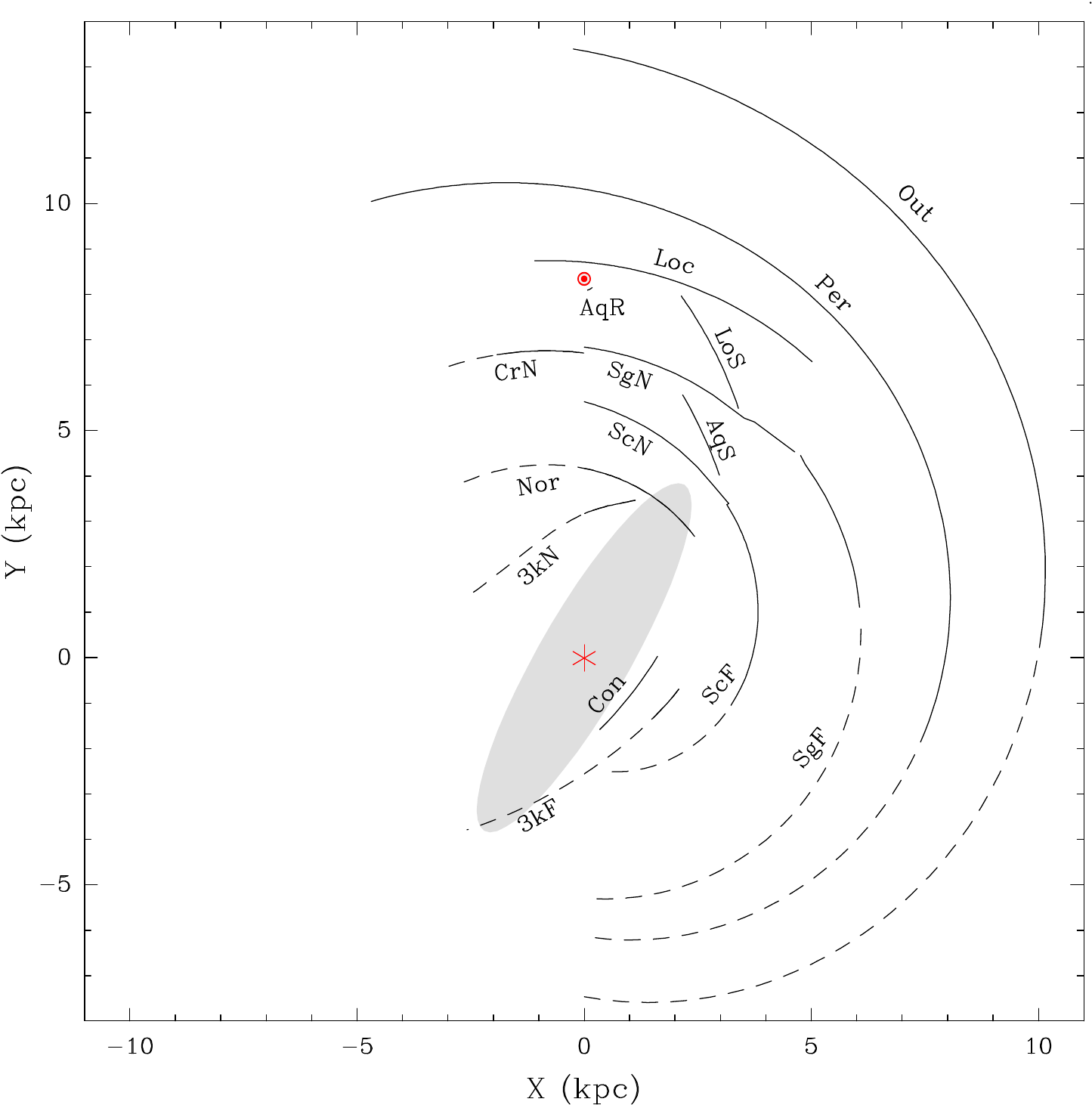}
\caption{\footnotesize
Locations and forms of spiral arm segments in the Milky Way as determined from
trigonometric parallaxes of water and methanal masers \citep[and references therein]{Reid:14} 
associated with HMSFRs ({\it solid lines}).  {\it Dashed lines} indicate extrapolations 
based on fits of log-periodic spiral forms to the parallax data.
Arm segments are labeled as follows: Outer (Out), Perseus (Per), Local (Loc),
Carina near portion (CrN), 
Sagittarius near and far portions (SgN, SgF), Scutum near and far portions (ScN, ScF), 
Norma or 4-kpc (Nor), 3-kpc near and far arms (3kN, 3kF), and Connecting (Con).  
Also shown are other features including the Aquila Rift (AqR), Aquila Spur (AqS),
and the Local spur (LoS).  The Outer Scutum-Centaurus (OSC), Carina far (CrF), an exension
of Connecting (CnX), and Centaurus-Crux near and far (CtN and CtF) arm segements currently 
have no parallax measurements and are not plotted.   
The Galactic Center and Sun are indicated with a {\it red asterisk} and {\it red Sun symbol}.
        }
\label{fig:arms}
\end{figure}

We used the fits of log-periodic spiral shapes to arm segments (slightly updated 
with some unpublished data) from the values summarized in \citet{Reid:14} to define the 
location of each arm segment (see Fig. \ref{fig:arms}).  For any given Galactic longitude, we
calculated the arm's Galactocentric radius and azimuth, $R$ and $\beta$, and the heliocentric
distance, $d$,  and combined this information with the arm \lbvA\ trace, to 
generate complete 6-parameter information, \lbvRBD, for all arm segments.  For the
Aquila Spur, Aquila Rift, and Connecting arm, distances were estimated from only one
or two parallaxes using simple linear forms.   The \lbvRBD\ information for 20
arm segments are currently used by the program, and we plan to update these and add
new arm segments when more parallax data is available.

We calculate a distance PDF, based on spiral arm information, as the 
product of the probability density of distance to arm segments, 
based on models fitted to parallax data and other information, 
and the probability of association with those arm segments:   
$$\Prob_{SA}(d|l,b,\vlsr,I) = \sum_{j=1}^J \Prob(d|\arm_j,l,b,\vlsr,I) \times \Prob(\arm_j|l,b,\vlsr,I)~~,$$
\noindent
where $I$ indicates prior information on the locations of $J$ spiral 
arm segments, as well as Galactic (\Ro, \To, rotation curve) and Solar Motion 
(\U, \V) parameters.   We assign the probability that a source resides in the $j^{th}$ arm 
segment as follows:
$$\Prob(\arm_j|l,b,\vlsr,I) = e^{-\Delta l_j^2/2\sigma_{l_j}^2}~
                              e^{-\Delta b_j^2/2\sigma_{b_j}^2}~
                              e^{-\Delta v_j^2/2\sigma_{v_j}^2}~~,$$

\noindent
where $\Delta l_j$, $\Delta b_j$, and $\Delta v_j$ are the minimum deviations in 
longitude, latitude, and velocity of the source from the center of the \lbvA\ trace
for the $j^{\rm th}$ spiral arm, respectively, and $\sigma_{l_j}$, $\sigma_{b_j}$, 
and $\sigma_{v_j}$ are their expected dispersions.  

The in-plane and out of plane dispersions are given by  
$\sigma_{l_j} = \sigma_s~\sec{\alpha}/d_{arm_j}$ and $\sigma_{b_j} = \sigma_z/d_{arm_j}$,
respectively, where $\sigma_s$ and $\sigma_z$ are the in-plane and out of plane widths
(Gaussian $1\sigma$) 
of an arm, and $\alpha$ is the angle between the tangent to the arm and a ray from the 
Sun through the target source.  The $\sec{\alpha}$ term accounts for the increased 
``effective'' width of the arm owing to its orientation with respect to its closest 
approach to the target ray.  Since real arms curve over kpc-scale lengths, we truncate 
$\sec{\alpha}$ above a value of 10.  

\begin{figure}[h]
\epsscale{0.85} 
\plotone{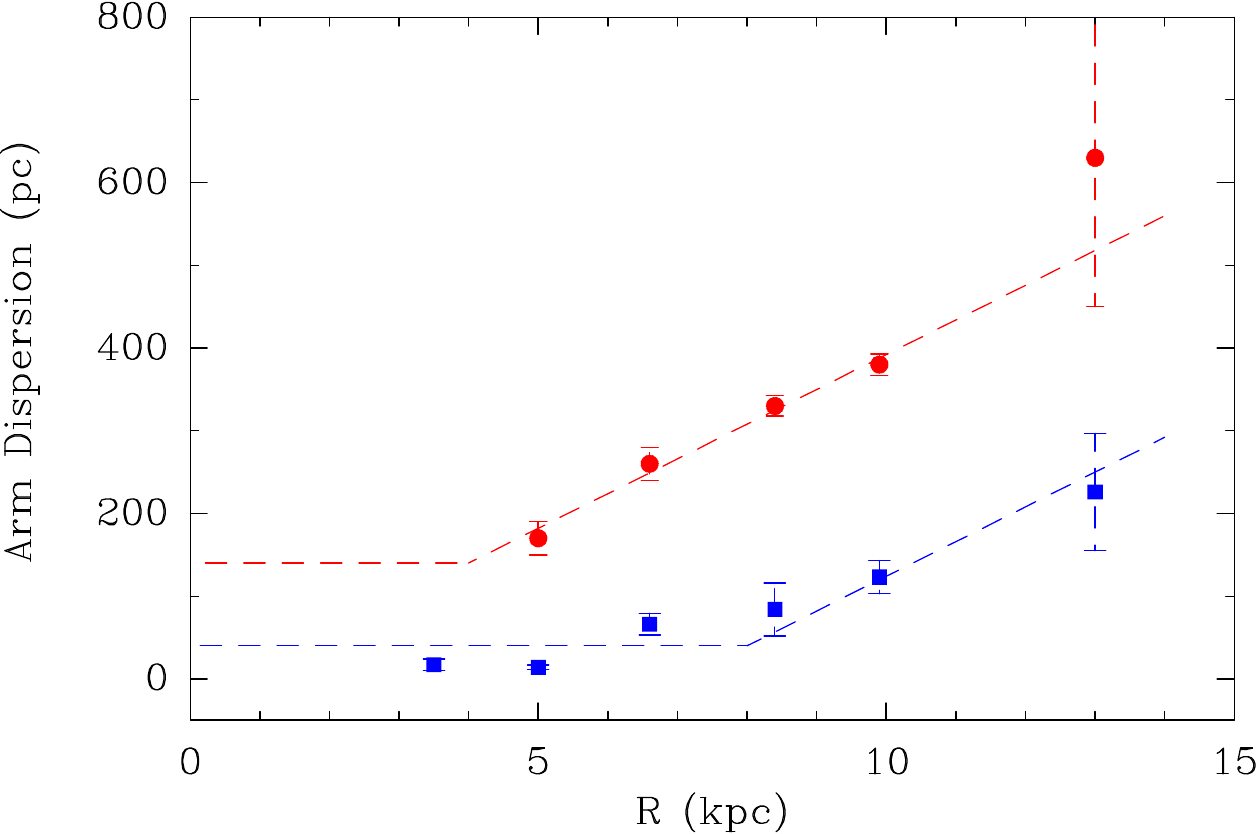}
\caption{\footnotesize
Spiral arm dispersion in width ({\it red circles}) and z-height ({\it blue squares})
versus Galactocentric radius after \citet{Reid:14}.   Data from parallax sources were
assigned to spiral arms and for each arm's data we calculated and plotted the 
standard deviation about the mean against the average radius.  
{\it Dashed lines} indicate broken linear fits, that were used to determine the 
expected deviations of sources from the model spiral arm segments.
        }
\label{fig:armwidths}
\end{figure}

Both $\sigma_s$ and $\sigma_z$ can be a function of Galactocentric distance.
\citet{Reid:14} found that spiral arm width increases with Galactocentric
radius, $R$, with a slope of 42 pc kpc$^{-1}$ for $R>5$ kpc.  Rather than 
extrapolate this to zero width near $R=0$, we adopt 
$\sigma_s=0.14 + 0.042(R-4)$~kpc for $R>4$ kpc and constant at $\sigma_s=0.14$ kpc 
in the central 4 kpc as shown in Fig.~\ref{fig:armwidths}.
In order to estimate the $z$-width, we grouped the same parallax data by spiral arm
and calculated the standard deviations about their mean offsets in $z$.   
These are also plotted in the figure.  The $z$-widths are consistent with increasing 
for $R>8$ kpc with the same slope as for the in-plane widths, and we adopt a simple 
broken-linear relation $\sigma_z = 0.04 + 0.042(R-8)$ kpc for $R>8$ kpc 
and 0.04 kpc inside of 8 kpc.
Finally, we set $\sigma_{v_j} = \sqrt2~\sigma_{\rm Vir}$, which is the 
expected Virial 1-dimensional speed {\it difference} for the target and a giant 
molecular cloud used to define the spiral arm \lbvA\ trace.  We adopt 
$\sigma_{\rm Vir}=5$ \kms, which is appropriate for $\sim10^6$ \Msun\ within 
$\sim100$ pc. 

In order to calculate $\Prob(d|\arm,l,b,\vlsr,I)$, we generate an array of
``trial'' source distances, $d$, and calculate the minimum separation 
of the source from a given arm segment model, both in the Galactic plane, $\Delta s$,
and out of the plane, $\Delta z$.   Then for the $j^{th}$ arm,
$$\Prob(d|\arm_j,l,b,\vlsr,I) \propto {1\over\sigma_s} e^{-\Delta s^2/2\sigma_s^2}
                          ~{1\over\sigma_z} e^{-\Delta z^2/2\sigma_z^2}~~,$$
\noindent
where the arm in-plane and z-widths are $\sigma_s$ and $\sigma_z$.
Note, we include the ``normalization'' terms, $1/\sigma_s$ and $1/\sigma_z$,
preceding the exponentials, since they vary with distance.  

Since the sum of the probabilities of association with arms 
$$S=\sum_{j=1}^{J} \Prob(\arm_j|l,b,\vlsr,I)$$
can be small, we add a uniform background probability that allows for the 
possibility of a missing arm segment in our model and
to some extent interarm sources (but see Section \ref{sect:spiral} for caveats).
A user-adjustable parameter, $P_{SA}$ from ${0\rightarrow1}$, determines the
maximum weighting of the spiral arm associations relative to the background probability. 
When normalizing the PDF, we set the total background probability (\ie\ integrated
over distance) to be $1- \min(S,P_{SA})$.  Thus, for example, if $S=0$, indicating
zero probability of the target being associated with spiral arms, then the
background probability would integrate to unity, regardless of the value 
assigned to $P_{SA}$.  However, if $S\ge1$, indicating a high probability
that the source can be assigned to one or more arms, then the spiral arm
PDF will be downweighted such that the background probability density would 
integrate to $1-P_{SA}$.   The program defaults to $P_{SA}=0.5$.

\subsection{Kinematic Distance PDF}\label{sect:KD}

The kinematic distance PDF, Prob$_{KD}(d)$, is generated for a given distance based 
on the velocity difference between the observed value and that expected from 
the rotation of the Galaxy.  First, we calculate a ``revised LSR velocity'' to allow 
for differences between the Standard Solar Motion (which defines LSR velocities) and more 
recent estimates.   We convert LSR to the Heliocentric velocities
by removing the IAU Standard Solar Motion of 20 \kms\ toward R.A.=$18^h$, Dec.=30\deg\ 
(in 1900 coordinates), which when precessed to J2000 give Cartesian Solar Motion 
components of (\Uo,\Vo,\Wo)=(10.27,15.32,7.74) \kms.  Then we calculate the
revised LSR velocity by applying (user-supplied) updated (\U,\V,\W) values.  Interestingly,
the IAU recommended values are surprisingly close to what could be considered today's
best values, and for simplicity the results in this paper adopt the IAU values.

Next, we obtain the velocity difference, \delV, between the revised LSR velocity and 
that expected from the rotation of the Galaxy.  Rather than use a simple linear 
rotation curve, we adopt the ``Universal'' rotation curve formulation of \citet{Persic:96} 
as implemented by \citet{Reid:14} (see their Figure 4 and Table 5 for details).  
This formulation is well motivated by observations of other galaxies, and 
it produces a nearly flat curve for Galactocentric radii $\gax5$ kpc
and slower rotation for sources inside that radius, in agreement with parallax 
and proper motion results.  

Assuming an individual star in a high-mass star forming region has a random 
(Virial) motion of $\sigV= 5$ \kms\ in one dimension, we calculate the distance PDF via
$$\Prob_{KD}(d|l,\vlsr,\sigV,I) \propto {1 - e^{-r^2/2}\over r^2/2}~~~~~~,$$
where $r = \delV/\sigV$.
We have used the \citet{Sivia:06} ``conservative formulation'' for the distance 
PDF, because it admits much higher probability for large deviations from the kinematic 
distance than would a Gaussian distribution.  This is critical as it better approximates 
the true probability density from kinematic information, which can have large systematic 
errors owing to significant peculiar motions.  See Section \ref{sect:examples} for
an example of an aberrant kinematic distance.  Note that by directly generating a
PDF, we have avoided calculating a traditional ``kinematic distance,'' which 
can have many complications when a velocity uncertainty is allowed.   

For some sources, one may have prior information on the probability that the
source is beyond the tangent point, near the far distance.   For example,
if \HI\ absorption is seen toward an \HII\ region with speeds exceeding that
of the region by $\gax20$ \kms, one can be fairly confident that the source
distance is past the tangent point.  (Although, there are places in the Galaxy 
where hydrogen gas has large non-circular motions that can lead to anomalous
absorption velocities.)   A user-supplied parameter, $P_{far}$, 
giving the prior probability (from 0 to 1) that the source is at the far distance, 
is used to weight the near and far PDFs before they are summed and normalized.  
When no such prior information is available, $P_{far} = 0.5$, and both near and 
far kinematic distance PDFs are given equal weight.

\subsection{Galactic Latitude PDF}

A source's Galactic latitude, $b$, is a direct indicator of distance, independent 
of spiral arm association, since the $\Prob(d|b)$ is taller and narrower for 
a more distant source than for a near one.  From Bayes' theorem, the distance PDF 
can be calculated from 
$$\Prob(d|b,\sigma_z,I)={\Prob(b|d,\sigma_z,I) \Prob(d|\sigma_z,I) \over 
                         \Prob(b|\sigma_z,I)}~~.$$
Neither $\Prob(d|\sigma_z,I)$ nor $\Prob(b|\sigma_z,I)$ introduce distance terms
and converting from latitude to $z$-height gives
$$\Prob(b|d,\sigma_z,I) \propto \Prob(z|d,\sigma_z,I)~{\partial z\over\partial b}~~.$$
Since $\partial z/\partial b = d$ and for a Gaussian distribution in $z$, we find  
$$\Prob(d|b,\sigma_z,I) \propto d~ e^{-z^2/2\sigma_z^2}~~.$$

The above formula is valid for a flat Galaxy.  Allowing for a Galactic warp, 
we replace $z$ in the above relation with the {\it difference}, $\Delta z$, 
between a source's Galactic height $z~(=d\times b)$ and a model of the Galaxy's warp.
Rather than adopt a mathematical formula for warping, we use the spiral arm \lbvRBD\ 
traces and calculate the $z$-height of each arm segment at the intersection of the 
arm with a ray from the Sun through a target source position.  
At any given distance from the Sun, we then estimate the 
$z$-height of the Galactic ``plane'' by interpolating (or extrapolating) linearly 
between ($d,z$) pairs of arm-warp parameters.  Typically, we obtain 2 or 3 such 
($d,z$)-pairs in the $2^{\rm nd}$ and $3^{\rm rd}$ quadrants and 4 to 6 pairs in 
the $1^{\rm st}$ quadrant.  In order to evaluate the PDF, we use the broken-linear
relation for $\sigma_z$ presented in Section \ref{sect:SA}

\subsection{Parallax Sources PDF}\label{sect:PS}

Sources with measured parallaxes have provided most of the information used to
fit segments of spirals that are included in our prior knowledge (I), although other 
information (such as tangent-point longitudes) can be included in the models.  
However, adopting a log-periodic spiral model for arm segments $\gax5$ kpc in length 
can be a simplification, and there is residual information in the 
parallax measures (\eg star formation substructure and deviation of the real arm from 
the mathematical model) which we would like to use.  So, we calculate 
a distance PDF based on association of the target source with a parallax source, 
Prob$_{PS}(d)$, as the product of the probability of distance given a parallax source, 
determined by its parallax accuracy, times the probability of association, 
given \lbv\ information for both the target and the parallax source.  
Summing over $K$ parallax sources gives
$$\Prob_{PS}(d|l,b,\vlsr,I) = \sum_{k=1}^K \Prob(d|PS_k,l,b,v,I) \times \Prob(PS_k|l,b,\vlsr,I)~~.$$

We quantify the association of a target source with a parallax source by calculating
the probability that they reside within the same giant molecular cloud (GMC), 
based on the separations in linear distance in longitude 
($\Delta d_l=d\Delta l$) and latitude ($\Delta d_b=d\Delta b$) and in velocity 
($\Delta\vlsr$):
$$\Prob(PS_k|l,b,\vlsr,I) \propto e^{-\Delta d_l^2/2\sigma_{GMC}^2} ~
      e^{-\Delta d_b^2/2\sigma_{GMC}^2} ~ e^{-\Delta\vlsr^2/2\sigma_{\Delta v}^2}~~,$$
\noindent
where we set the expected 1-dimensional separation of two sources within a GMC to be
$\sigma_{GMC}=0.05$ kpc and the expected separation in line-of-sight velocity to be 
$\sigma_{\Delta v} = \sqrt{2}~\sigma_{\rm Vir} \approx 7$ \kms.

For the $k^{\rm th}$ source with measured parallax, $p \pm \sigma_p$, we calculate
$$\Prob(d|PS_k,I) \propto p^2~e^{-\Delta p^2/2\sigma^2_p}~~,$$
where for a given distance $d$, $\Delta p = (1/d) - p$.  Note, since what is measured
is parallax, not distance, we have employed $\Prob(d) = \Prob(p) \times p^2$ in the PDF 
calculation.  Owing to the possibility of multiple parallax sources 
(possibly from different spiral arms) having similar \lbv\ values as the target source, 
we limit the impact of this probability term by adding a constant background probability 
density to $\Prob_{PS}(d|l,b,\vlsr,I)$ based on the cumulative weight of the parallax 
matches: 
$$W_\pi = \sum_{k=1}^K \Prob(PS_k|l,b,\vlsr,I)~~.$$  
A user-adjustable parameter, $P_{PS}$ from ${0\rightarrow1}$, determines the
maximum weighting of the parallax source association relative to the background probability. 
When normalizing the PDF, we set the total background probability integrated
over distance to be $1- \min(W_\pi,P_{PS})$.  (See the discussion in Section \ref{sect:SA}.)
In light of the partial correlation of individual parallax source information with that
in the spiral arm model, the program defaults to a moderately-low weighting of this
information by setting $P_{PS}=0.25$.

\section{Example Distance PDFs} \label{sect:examples}

\begin{figure}[h]
\epsscale{0.75} 
\plotone{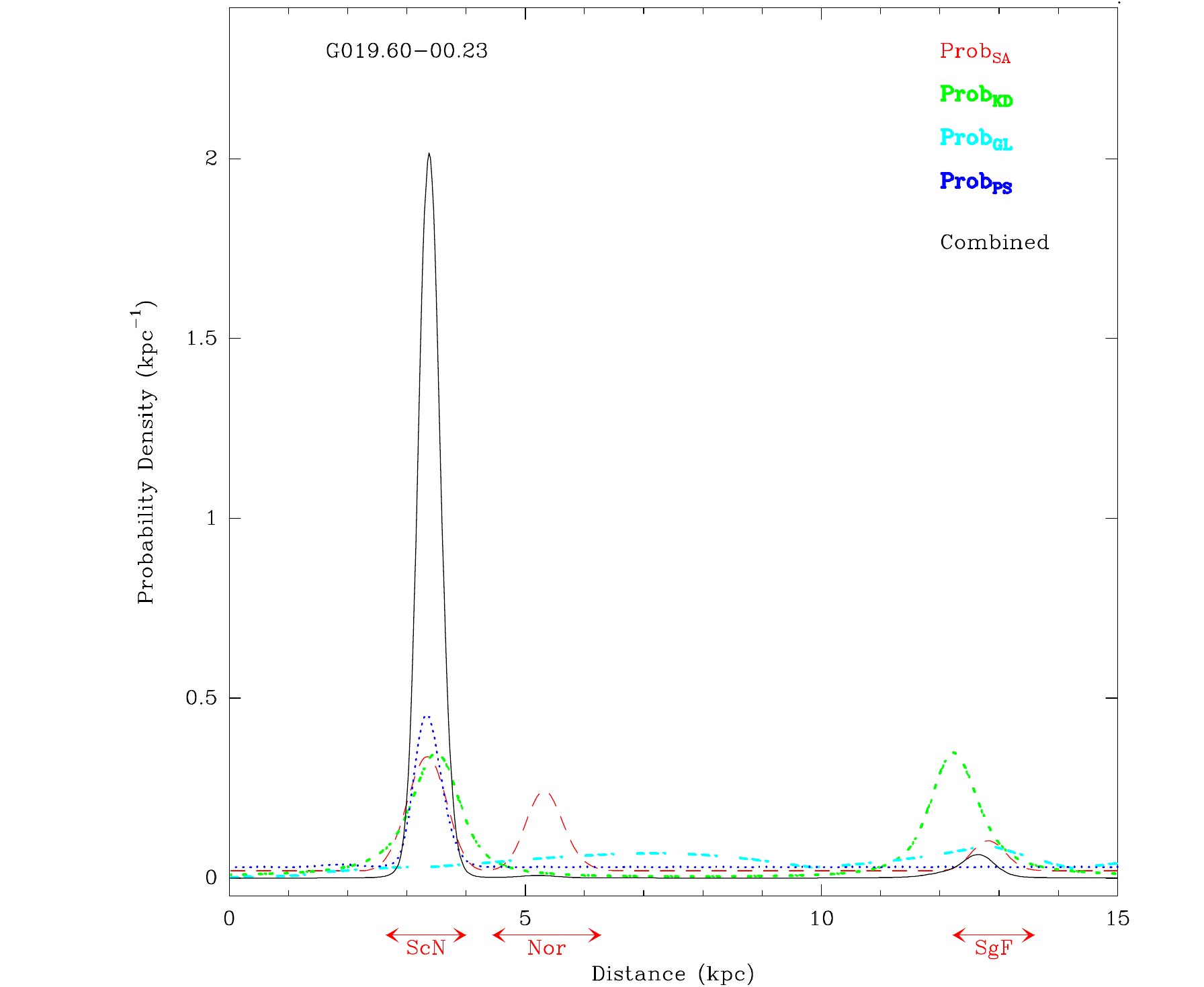}
\caption{\footnotesize
Distance probability density function for G019.60$-$00.23, a 22 GHz water maser at 
$v = 47$ \kms\ from the \citet{Valdettaro:01} catalog.  The \lbvS\ values most likely 
associate it with the near portion of the Scutum spiral arm (at 3.3 kpc), although
the Norma arm (at 5.4 kpc) and the far portion of the Carina-Sagittarius arm (at 12.9 kpc) 
are not ruled out ({\it red dashed lines}).  
However, the combination ({\it solid black line}) 
of arm assignment probability with the kinematic distance ({\it green dot-dashed line}), 
latitude probability ({\it cyan dashed line}) and association with a parallax source 
({\it blue dotted line}) strongly favor a distance of $3.38\pm0.19$ kpc.
        }
\label{fig:G019}
\end{figure}

We present examples of the distance PDFs for sources in the $1^{st}$ and $2^{nd}$
Galactic quadrants. G019.60$-$00.23 is a star formation water maser at $\vlsr=47$ \kms\ 
from the \citet{Valdettaro:01} catalog and its distance PDF is shown in Fig.~\ref{fig:G019}.
This source was found to have a high probability of association with the near
portion of the Scutum arm at a distance of 3.3 kpc and lower probabilities of association 
with the Norma arm at 5.3 kpc and the far portion of the Sagittarius arm at 12.9 kpc.  
Both the near kinematic distance and the association with a parallax source 
(G018.87+00.05) strongly favor the near Scutum arm distance.  Combining all information, 
G019.60$-$00.23 is estimated to be at a distance of $3.38\pm0.18$ kpc in the near portion 
of the Scutum arm with an integrated probability of 95\% (and a 5\% probability at 
$12.62\pm0.32$ kpc in the far portion of the Sagittarius arm).

\begin{figure}[h]
\epsscale{0.75} 
\plotone{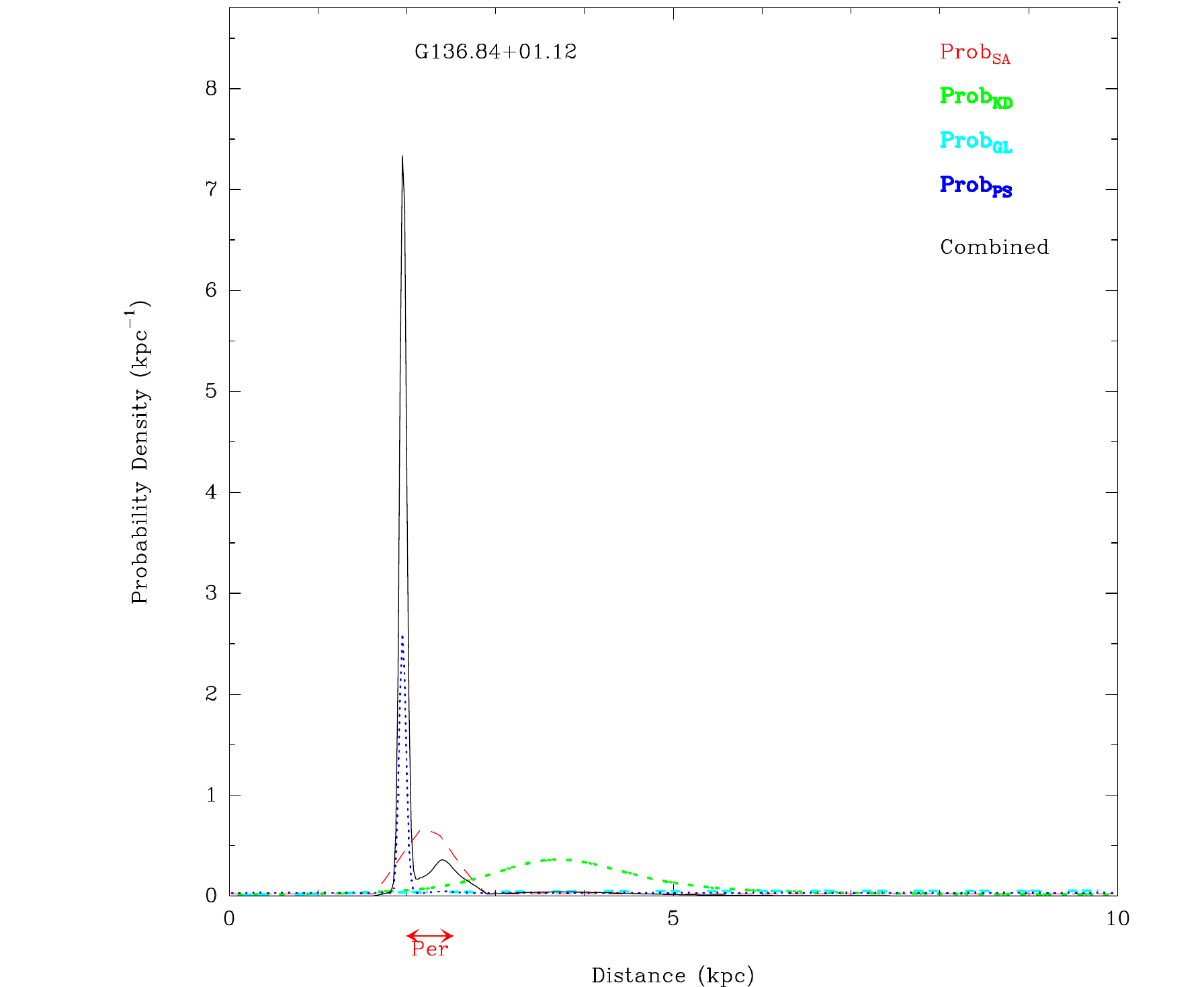}
\caption{\footnotesize
Distance PDF for G136.84+01.12, a 6.7 GHz methanol maser at $\vlsr = -45$ \kms\ 
from the \citet{Pestalozzi:05} catalog.  
The \lbvS\ values associate it with the Perseus spiral arm centered
near 2.4 kpc ({\it red dashed line}).  This information, coupled with association
with the parallax source G133.95+1.05 ({\it blue dotted line}), favors the 
distance estimate of $1.96\pm0.04$ kpc over the arm center.  Note that the kinematic 
distance ({\it green dot-dashed line}) suggests a much greater distance of 3.7 kpc,
as it does for many sources in this part of the Perseus arm which is known to
be kinematically anomalous.
        }
\label{fig:G136}
\end{figure}

G136.84+01.12 is a 6.7 GHz methanol maser source at $\vlsr = -45$ \kms\ in the 
\citet{Pestalozzi:05} catalog, and its distance PDF is shown in Fig.~\ref{fig:G136}.   
This source has a very high probability of association with the Perseus spiral arm 
(which crosses the source's longitude at a distance of $\approx2.4$ kpc).   
It may be associated with a giant molecular cloud containing the parallax source 
(G133.95+1.05) near the inner edge of the arm.   The kinematic distance for 
G136.84+01.12 is 3.7 kpc, which, if correct, would place it in the Outer spiral arm.  
However, this portion of the Perseus arm is well known for kinematic anomalies, and the 
program correctly recognizes that the source \lbvS\ values are inconsistent with 
Outer arm values.  This is an example
of the utility of a non-Gaussian PDF (see Section \ref{sect:distance}) for kinematic
distances, as it accommodates such anomalies.  The program favors a distance 
of $1.96\pm0.04$ kpc (based on the very accurate parallax to a nearby source) 
over a slightly greater distance of $2.36\pm0.27$ kpc (from the arm mid-line), 
as is apparent from an examination of the PDF.

\section{Comparisons with Other Distance Estimates} \label{sect:comparisons}

A common method to estimate distances to Galactic sources of star formation
is to use kinematic distances, augmented by \HI\ absorption spectra to
resolve the near/far ambiguity for sources within the Solar circle.  
Recently, \citet{Anderson:12} discovered large numbers of \HII\ regions
in the \firstQ\ quadrant, and for most they resolved the distance ambiguity
via \HI\ absorption.  They found 62 sources that could be confidently 
(``A-grade'') placed at the far kinematic distance.   Using our Bayesian approach 
with no prior information to resolve the near/far ambiguity (\ie $P_{far}=0.5$),
we were able to assign distances with 90\% or greater probability for 34 of 
these sources.  Of these sources, there was agreement between the \HI\ absorption 
and our technique as to the resolution of the near/far distance 
ambiguity for all but six sources.

The techniques disagree for two sources near 19\deg\ longitude.  For 
G018.324+0.026 and G019.662$-$0.305, the Bayesian approach finds a significantly 
more probable association with the near portion of the Scutum arm at a distance 
of 3.4 kpc than with the far portion of the Sagittarius arm at 13 kpc.   
Similarly, we find G029.007+0.076 is likely associated 
with the near portion of the Scutum arm at a distance of $\approx4$ kpc, 
whereas at the far kinematic distances of $\approx11$ kpc there are no probable 
arm associations.  If these sources are at the distances suggested by
our Bayesian program, then they are near the end of the Galactic bar and may
have significant non-circular motions, which could confuse distances based 
on kinematics and \HI\ absorption.   

For the remaining three sources (G032.272$-$0.226, G034.041+0.053 and G034.133+0.471) 
we find them likely associated with the near portion of the Sagittarius arm
at $\approx2.1$ kpc rather than at the far kinematic distance of $\approx12$ kpc.
Note that along our line of sight toward longitude $\approx34$\deg\ there
may be gas with anomalous velocities owing to interaction with the supernova
remnant W~44 (centered at ($l,b$) = ($34\d69,-0\d41$) with a radius of 0\d6) 
or with winds from its precursor.

Even though our program suggests a near distance for these six sources with greater 
than 90\% probability, that still leaves a non-negligible probability for an 
alternative distance.  All in all, we conclude that when a source can be assigned 
to a spiral arm with high probability, there is generally good agreement between 
our Bayesian approach and using \HI\ absorption to resolve the near/far distance 
ambiguity.

\section{The Milky Way's Spiral Structure} \label{sect:spiral}

\begin{figure}[ht]
\epsscale{1.0} 
\plottwo{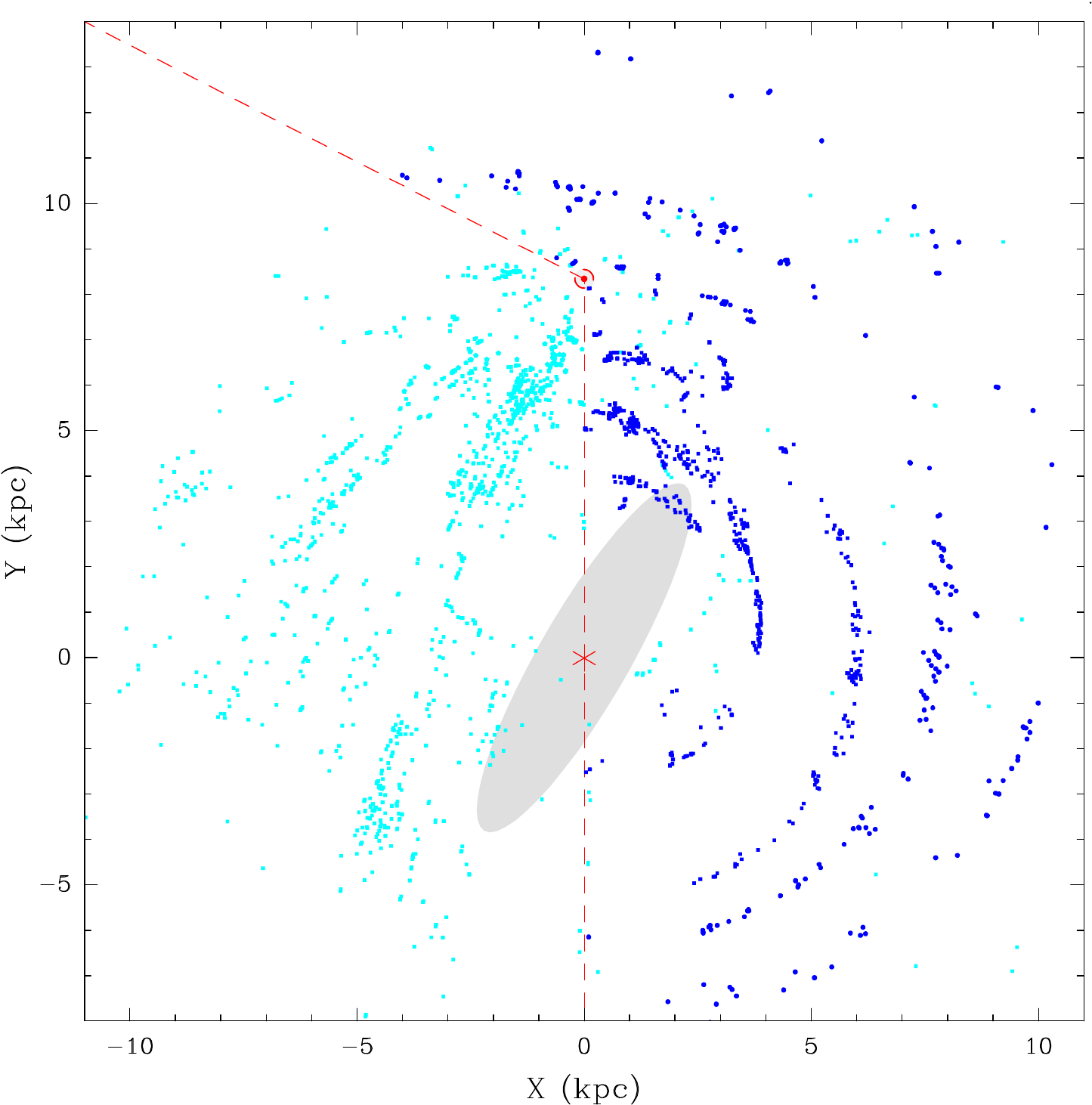}{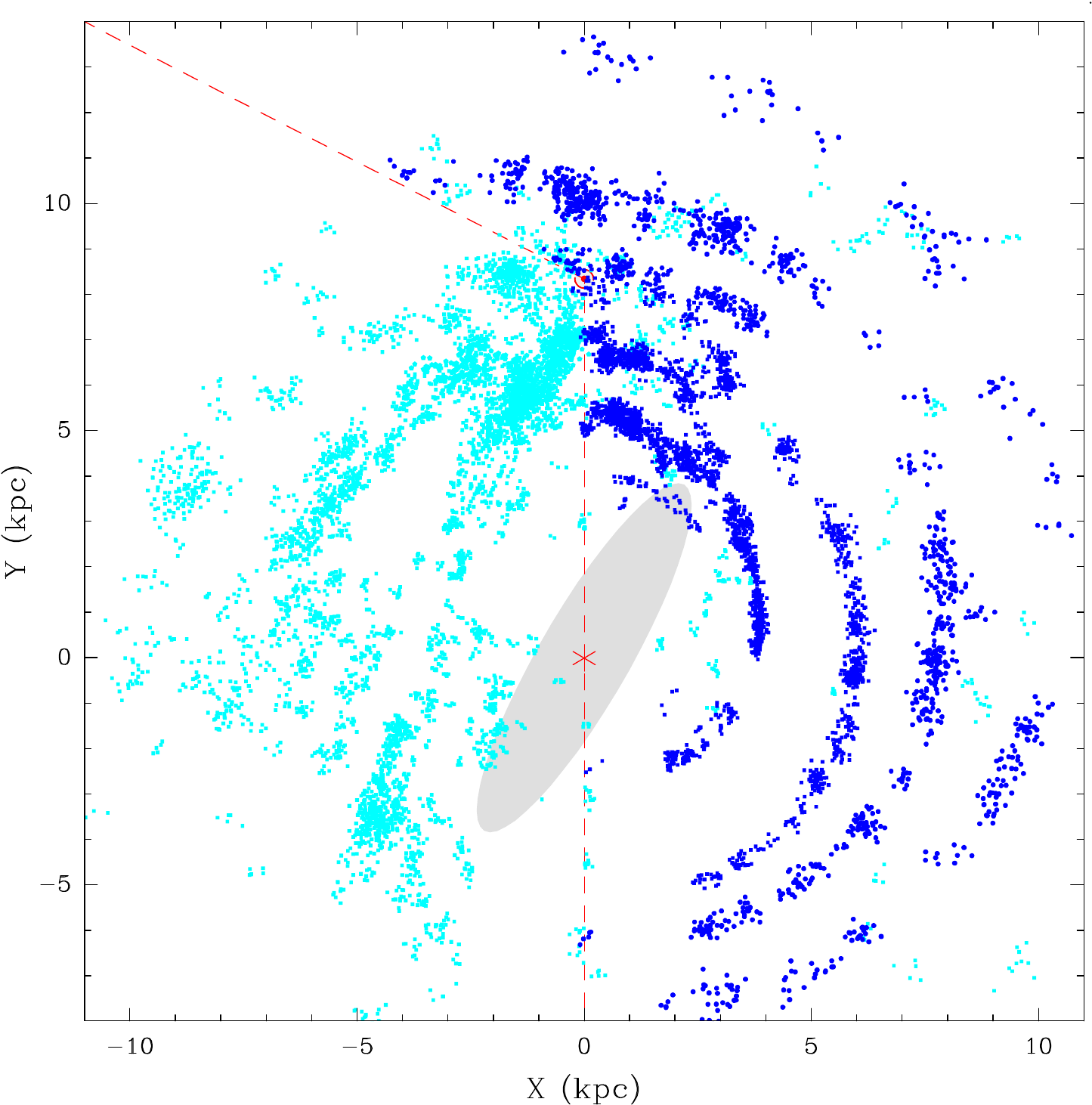}
\caption{\footnotesize
{\it Left Panel:} Map of high-mass star forming regions that trace spiral structure in 
the Milky Way viewed from the North Galactic Pole.  
Distances were determined with the Bayesian approach 
described in this paper, which combines spiral arm locations, association 
with giant molecular clouds that include sources with measured parallaxes, 
kinematic distance and Galactic latitude information.  
Input catalogs included water and methanol masers, HII regions, and ``red'' MSX sources.
{\it Dark blue dots} indicate sources associated with spiral arms, while the 
lighter {\it cyan dots} could not be confidently associated with any arm.  
The {\it dashed red lines} separate the regions of the Milky Way mapped with
northern hemisphere telescopes ({\it right side}) from that yet to be mapped 
with southern hemisphere telescopes ({\it left side}).
The shaded ellipse provides a schematic indication of the Galactic bar.
{\it Right Panel:} Same as the left panel, except that
each catalog source is represented by 5 ``dots,'' spread randomly about the estimated
position using a Gaussian kernel with $\sigma=100$ pc to better represent
multiple sources of star formation in a giant molecular cloud.  Spiral arms evident
in the figures, starting from near the Galactic center and moving outward, are
the Norma, Scutum, Sagittarius, Local, Perseus, and Outer arms. 
        }
\label{fig:map}
\end{figure}

One application of our Bayesian distance estimation program is to generate a ``plan 
view'' (\ie\ a projected view from high above the plane) of the spiral structure of 
the Milky Way.  The left panel of Fig.~\ref{fig:map} 
shows the locations of $\approx2000$ HMSFRs, assembled by combining catalogs of water 
\citep{Valdettaro:01} and methanol \citep{Pestalozzi:05} masers, HII regions 
\citep{Anderson:12}, and ``red'' MSX sources \citep{Urquhart:14} sources.  
The distance to each source was determined from the component fitted to the combined 
distance PDF that had the greatest integrated probability. 
On the right-hand portion of this figure, where spiral arm locations are established by
parallaxes, approximately 90\% of the catalog sources are found to be associated with
a spiral arm (indicated by the dark blue dots); the remaining approximately 10\% 
probably represent interarm star formation (indicated by the light cyan dots). 
Since we do not yet have sufficient parallax measurements to locate spiral arms in the 
southern hemisphere (left-hand portion this figure), distance estimation there is
based only on kinematics and latitude.  The limitations of kinematic distances, including
being multi-valued in the 4$^{th}$ quadrant and having variable sensitivity to non-circular 
motion, blur the spiral structure in this part of the Milky Way.

In the right panel of Fig.~\ref{fig:map} we present a visualization of the Milky Way. 
For each catalog source we ``sprinkle'' five dots to simulate multiple sites of
star formation in each giant molecular cloud.   These dots were shifted from the
catalog source location by adding Gaussianly distributed shifts with $\sigma=100$ pc
along each axis.  The visualization suggests a clumpy spiral structure 
with depressions or gaps in massive star forming regions.  The dearth of points in the 
Perseus arm between Galactic longitudes of 50\deg\ to 80\deg, pointed out by 
\citet{Zhang:13} based on the lack of parallax sources, is apparent, even though the Bayesian
distance estimation has no bias against portions of a spiral arm.  
The Outer and Sagittarius arms also show two or more gaps, several kpc in length.   

The assumptions that go into, and the limitations of, Fig.~\ref{fig:map} deserve
discussion.  Firstly, the accuracy of this figure directly depends on the locations of 
spiral arms, based on parallax results.  Currently, the location of the Outer arm at 
Galactic longitudes $\lax70\deg$, the Perseus and Sagittarius arms at longitudes 
$\lax40\deg$, and the Scutum arms at longitudes $\lax25\deg$ are extrapolations 
based on pitch angles fitted to parallax sources at greater longitudes.  
Since spiral arms may have pitch angles that vary with azimuth \citep{Honig:15}, 
distances to sources in the extrapolated longitude ranges should be considered tentative,
pending more parallax data.

\begin{figure}[ht]
\epsscale{0.65} 
\plotone{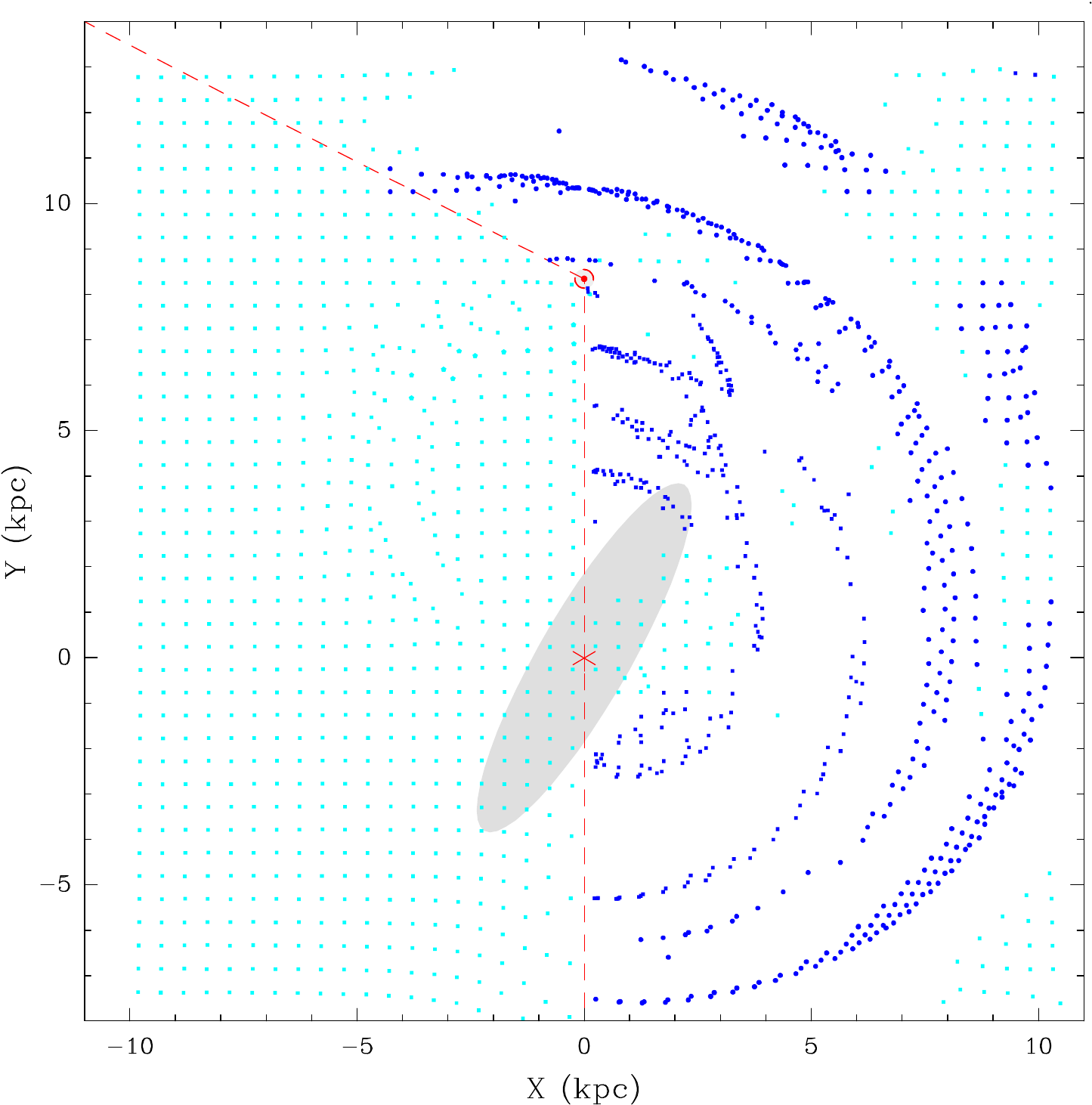}
\caption{\footnotesize
Test of the influence of including the spiral arm contribution to the 
combined distance PDF.  We generated fake sources on a uniform grid with 
pure circular rotation.  For sources to the left and below of the {\it red dashed lines},
where no spiral arm information is used, the program returns the input locations with
reasonable accuracy  (except for slight deviations in regions near $\ell\approx0$ and
near the ``tangent points'' where small velocity changes result in large
changes in distance).  However, to the right and above the {\it red dashed lines},
spiral arm information was used, resulting in distance estimates biased to
spiral arms.
        }
\label{fig:grid}
\end{figure}

Secondly, even though we include a flat (non-zero) background probability density 
in the spiral arm PDF (see Section~\ref{sect:SA} for details), including 
$\Prob_{SA}(d)$ in the combined distance PDF will ``pull'' peaks of 
combined probability density toward an arm center.  To demonstrate this, we
generated a uniform grid of sources that follow circular Galactic orbits, 
and estimate distances to these test sources.  In order to simplify this 
demonstration, the test sources were placed exactly in the Galactic plane
and the Galactic latitude, $\Prob_{GL}(d)$, and parallax source, $\Prob_{PS}(d)$,
contributions to the combined distance PDF were ignored.  The results of this
uniform grid test are shown in Fig.~\ref{fig:grid}.  In regions
of this uniform mock ``Galaxy,'' where spiral arm information is lacking, the grid is
reasonably well preserved.  However, where the spiral arm information exists 
and is used in the combined distance PDF, the results are biased to arms.
This highlights the effects of the assumption made in the Bayesian distance
estimation method that sources whose \lbvS\ values are similar to the \lbvA\ values of a
known spiral arm segment probably are in that arm.   Therefore, one should not
use this method for sources that are not, or only weakly, associated with spiral 
structure, such as populations of old stars or diffuse gas.
 
\vskip 0.5truein \noindent 
This work was partially funded by the ERC Advanced Investigator Grant GLOSTAR (247078).

\vskip 0.5truein\noindent 
{\it Facilities:}  \facility{VLBA}, \facility{VERA}, \facility{EVN}

\section{Appendix}\label{sect:appendix}

\begin{figure}[ht]
\epsscale{0.85} 
\plotone{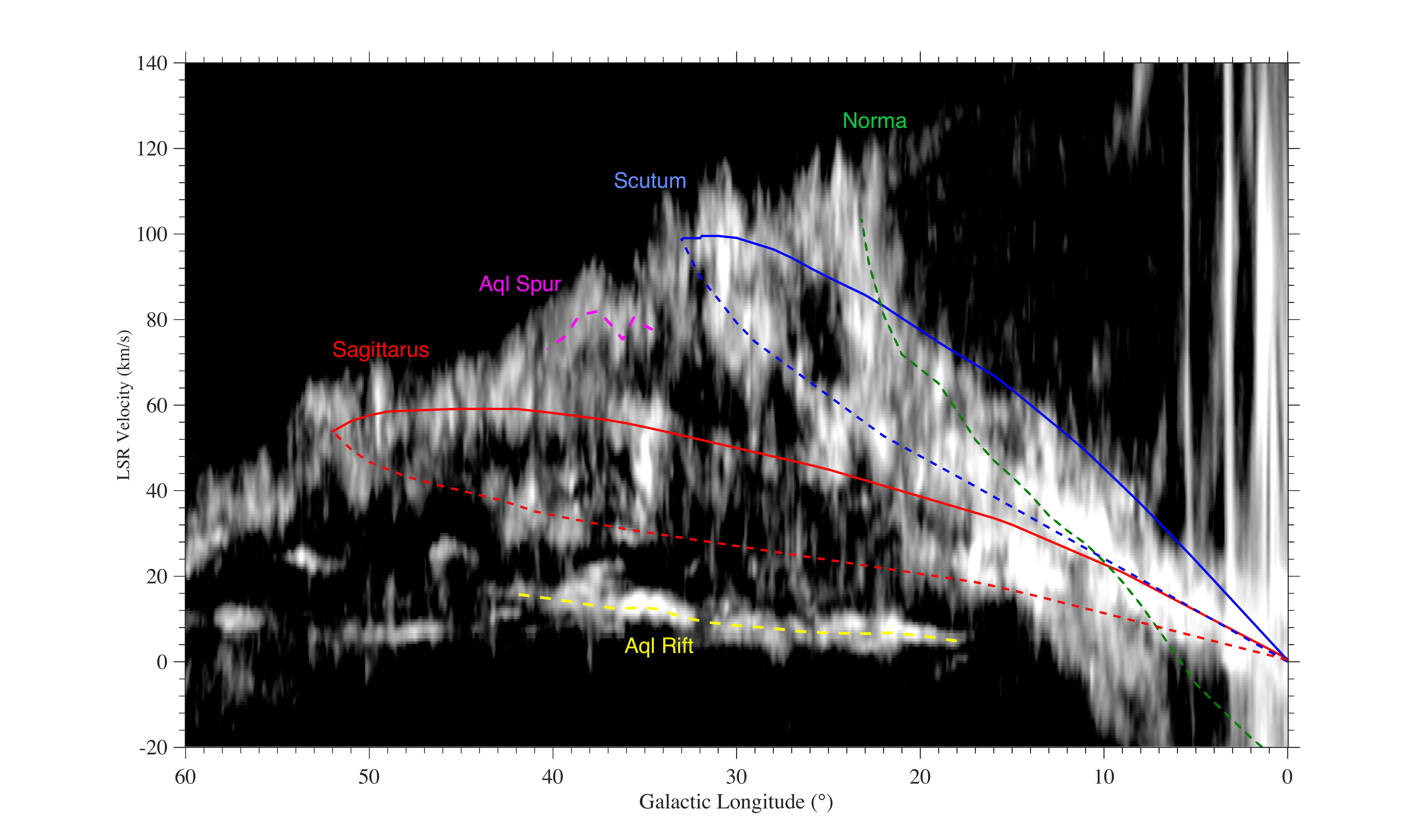}
\caption{\footnotesize
Traces of the Sagittarius, Scutum, and Norma arms in the first Galactic quadrant, 
as well as the smaller Aquila spur \citep{Cohen:80,Dame:86} and Aquila 
Rift features, on the CfA CO survey integrated from $b = -1\deg~{\rm to}~+1\deg$. 
The near and far sides of the arms are {\it dotted} and {\it solid} lines, respectively. 
The far side of Scutum probably does not extend below $l \sim 20\deg$. 
The near side of the Sagittarius arm passes so close to the Sun that its tracing clouds are 
widely separated in longitude.
        }
\label{fig:quad1}
\end{figure}

\begin{figure}[ht]
\epsscale{0.85} 
\plotone{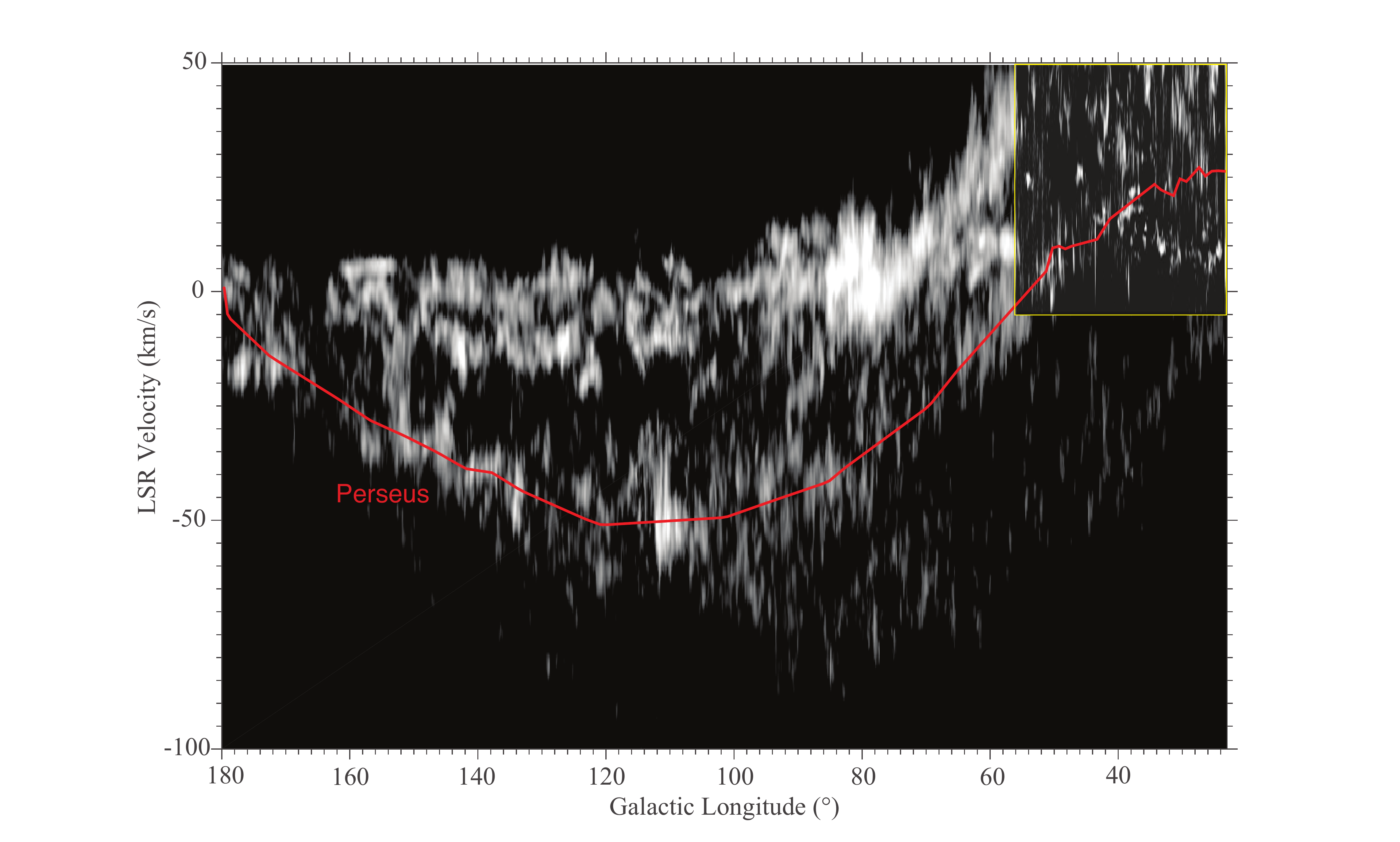}
\caption{\footnotesize
A trace of the Perseus arm in the first and second quadrants.  Over the longitude range 
$l= 56\deg~{\rm to}~180\deg$ the trace follows the CfA CO survey integrated from 
$b = -2\deg~{\rm to}~+2\deg$.  At lower longitudes (within the yellow box) the arm 
spirals into the far side of the inner Galaxy, where its large distance ($\approx12$ kpc) 
and confusion with near-side emission at the same velocity make the arm difficult to follow. 
In that region, the higher-resolution CO Galactic Ring Survey \citep{Jackson:06} 
is integrated over a half-degree strip of latitude that roughly follows the arm in latitude 
as judged by visual examination of 21-cm latitude-velocity maps from the VGPS survey 
\citep{Stil:06}.
        }
\label{fig:q1q2_Perseus}
\end{figure}

\begin{figure}[ht]
\epsscale{0.85} 
\plotone{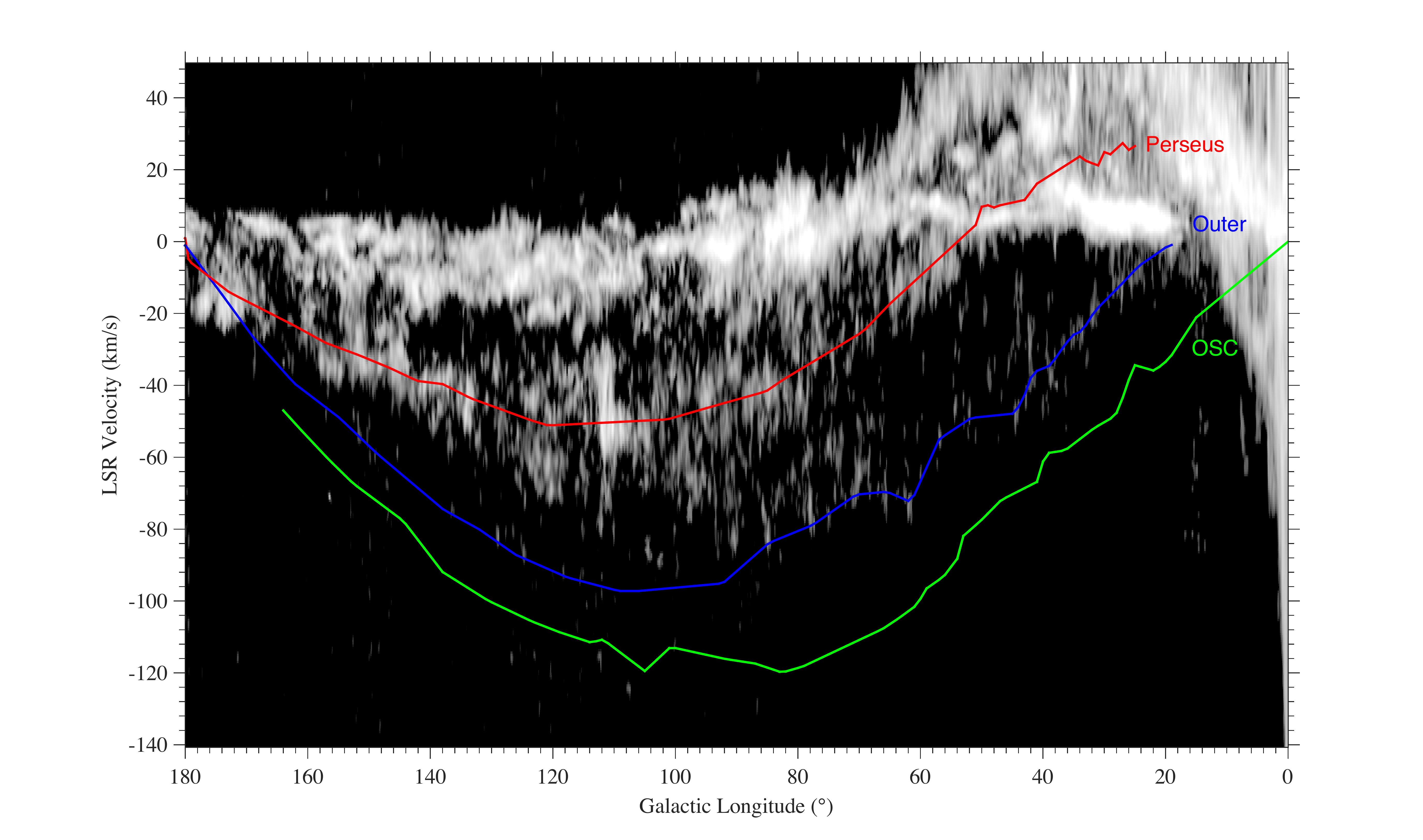}
\caption{\footnotesize
A trace of the Outer arm on the CfA CO survey integrated from $b = -5\deg~{\rm to}~+5\deg$.  
The survey was moment masked to suppress noise while integrating over such a large range of 
latitude. The traces of the Perseus and Outer Scutum-Centaurus arms are also shown for 
reference; note that the CO map here does not show those two arms optimally 
(see instead Figs.~\ref{fig:q1q2_Perseus} \& \ref{fig:q1q2_OSC}). 
The emission near $-70$~\kms\ between $l = 65\deg~{\rm and}~100\deg$ may be a bridge 
between the Perseus and Outer arms.
        }
\label{fig:q1q2_Outer}
\end{figure}

\begin{figure}[ht]
\epsscale{0.85} 
\plotone{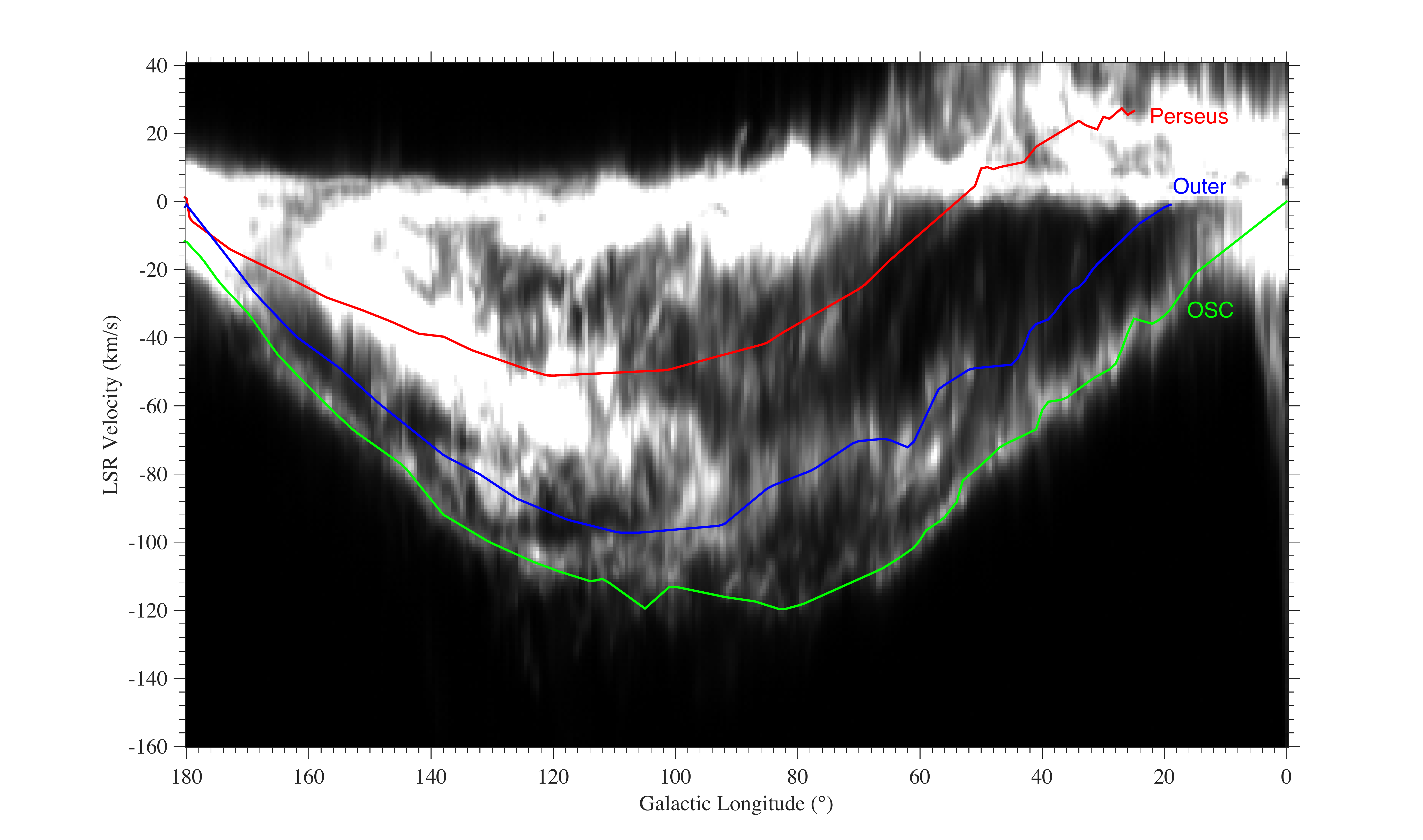}
\caption{\footnotesize
A trace of the Outer Scutum-Centaurus arm on the Leiden-Argentine-Bonn 
21cm survey integrated over a $1\deg$ 
strip of latitude that roughly follows the arm.  The traces of the Outer and Perseus arms 
are also shown for reference; note that this 21-cm map does not show those two arms optimally.
        }
\label{fig:q1q2_OSC}
\end{figure}

\begin{figure}[ht]
\epsscale{0.85} 
\plotone{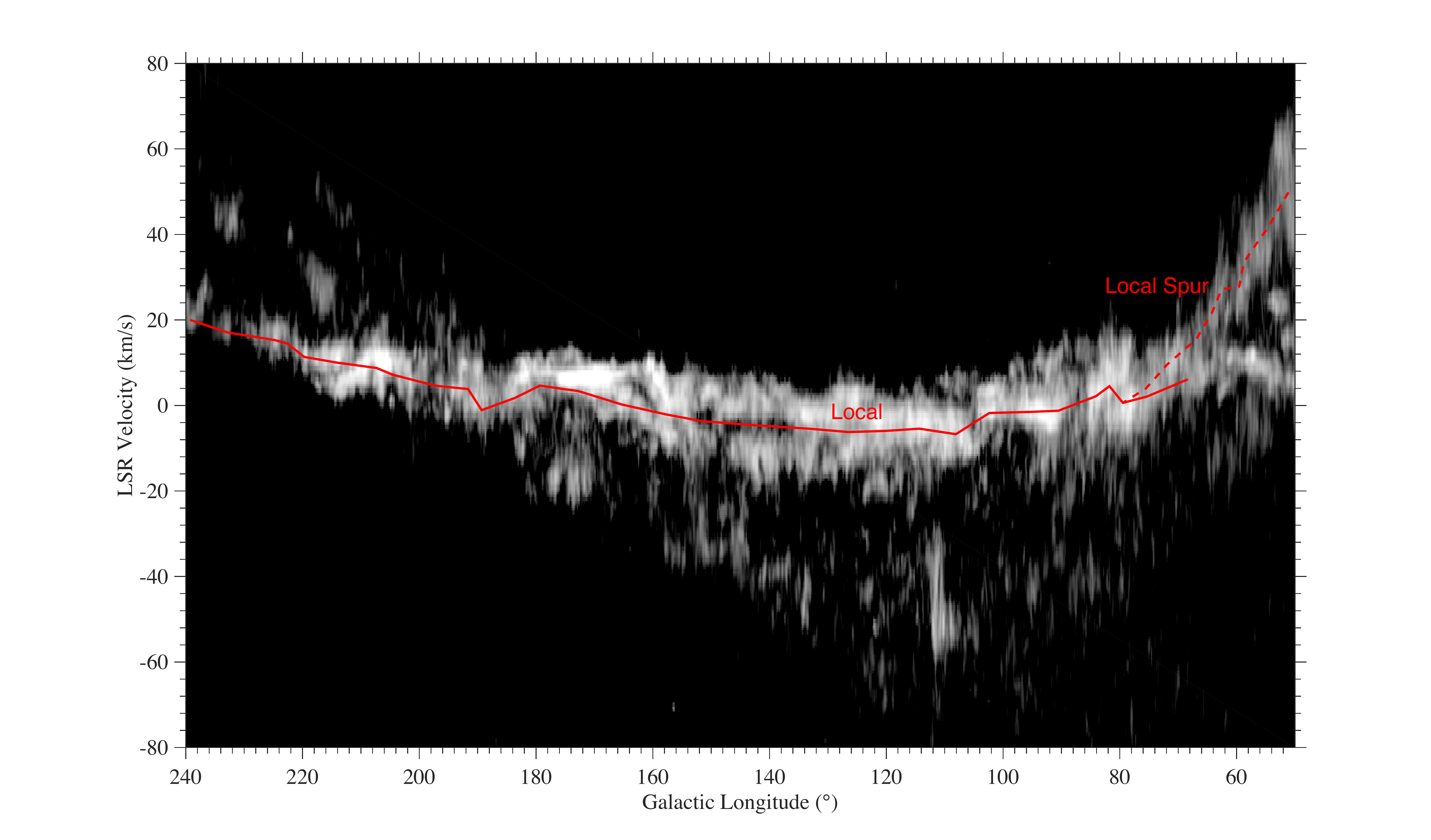}
\caption{\footnotesize
Traces of the Local arm and a high-inclination spur that apparently stretches from the 
Local arm to the region of the Sagittarius arm tangent point (near $l=50\deg,v=60$ km/s). 
The spur is traced by about six high mass star forming regions with maser parallax distances 
(Xu et al. in preparation) as well as by CO.  The underlying grayscale is the CfA CO survey 
integrated from $b=-30\deg~{\rm to}~+30\deg$; the wide latitude range emphasizes nearby material. 
Both traces closely follow the \lv\ loci defined in Xu et al.
        }
\label{fig:local}
\end{figure}

\begin{figure}[ht]
\epsscale{0.85} 
\plotone{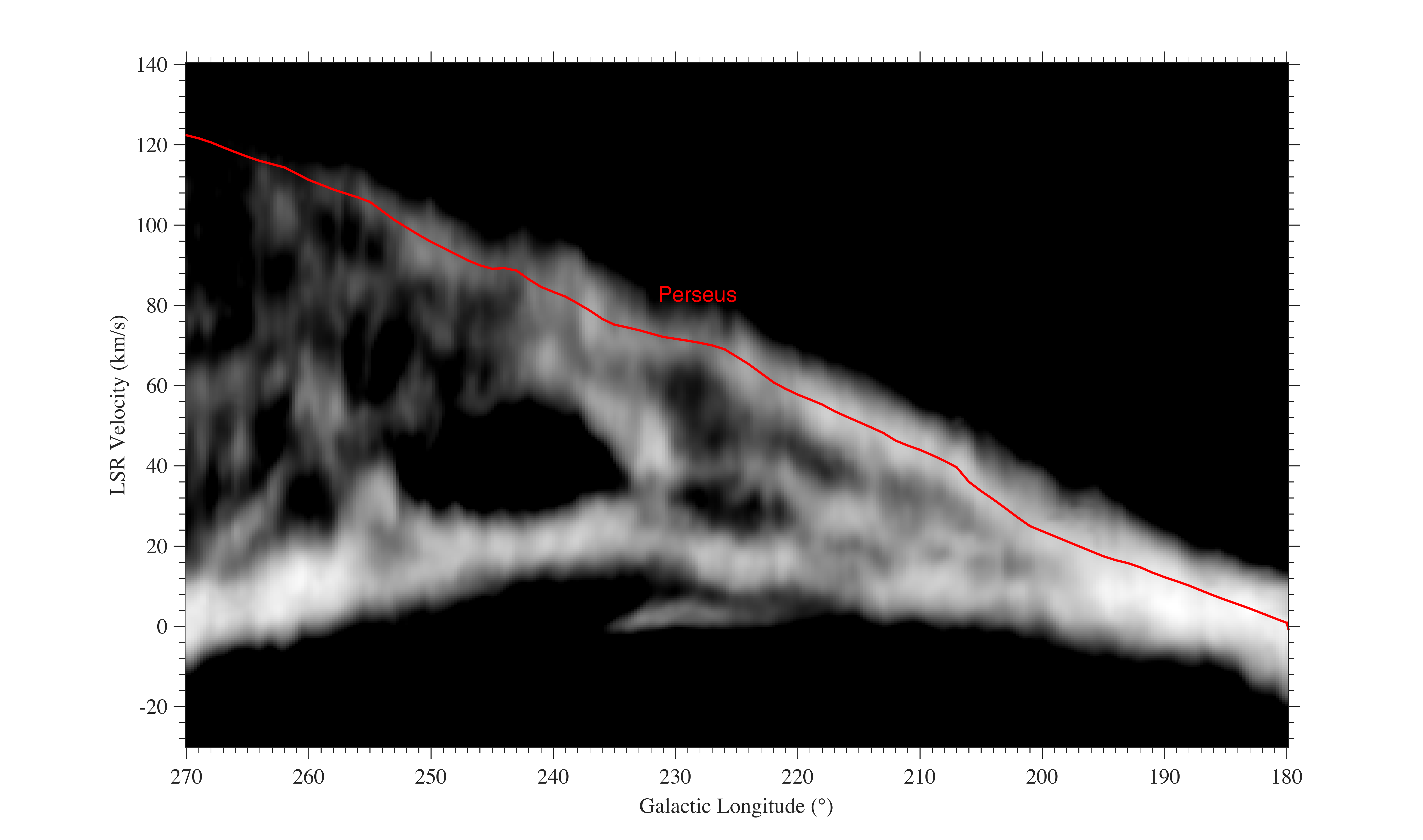}
\caption{\footnotesize
A trace of the Perseus arm in the third Galactic quadrant on the Leiden-Argentine-Bonn 21-cm 
survey integrated from $b = -5\deg~{\rm to}~+2\deg$.
        }
\label{fig:quad3}
\end{figure}

\begin{figure}[ht]
\epsscale{0.85} 
\plotone{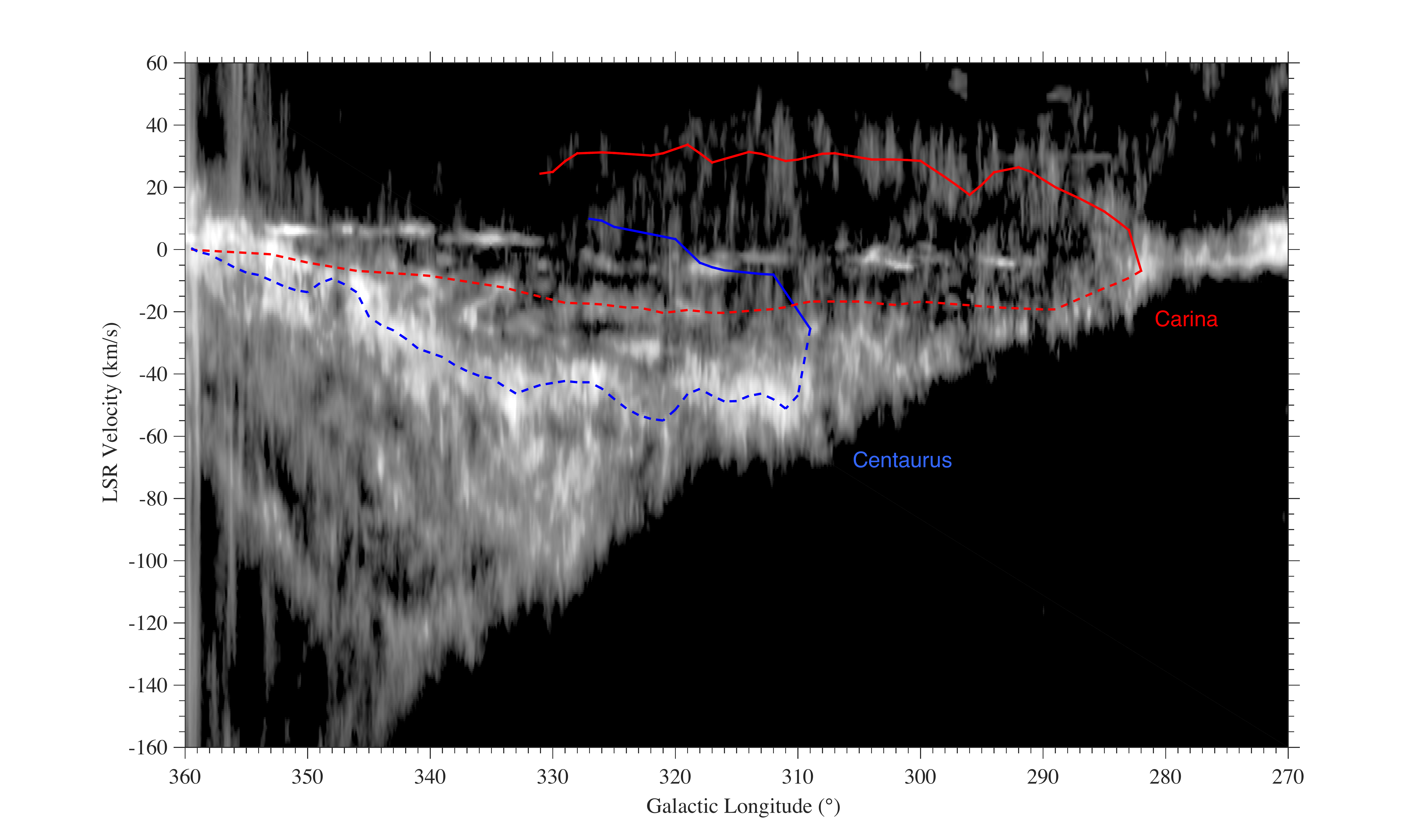}
\caption{\footnotesize
Traces of the Centaurus and Carina arms on the CfA CO survey integrated from 
$b = -5\deg~{\rm to}~+5\deg$.  The survey was moment masked to suppress noise while integrating 
over such a large range of latitude.  The near and far sides of the arms are {\it dotted} and 
{\it solid} lines, respectively.  The near side of Carina passes so close to the Sun that its 
tracing clouds are widely separated in longitude. The far side of Centaurus arm is difficult to 
trace owing to its great distance ($\sim14$~kpc).
        }
\label{fig:quad4}
\end{figure}

\begin{figure}[ht]
\epsscale{0.85} 
\plotone{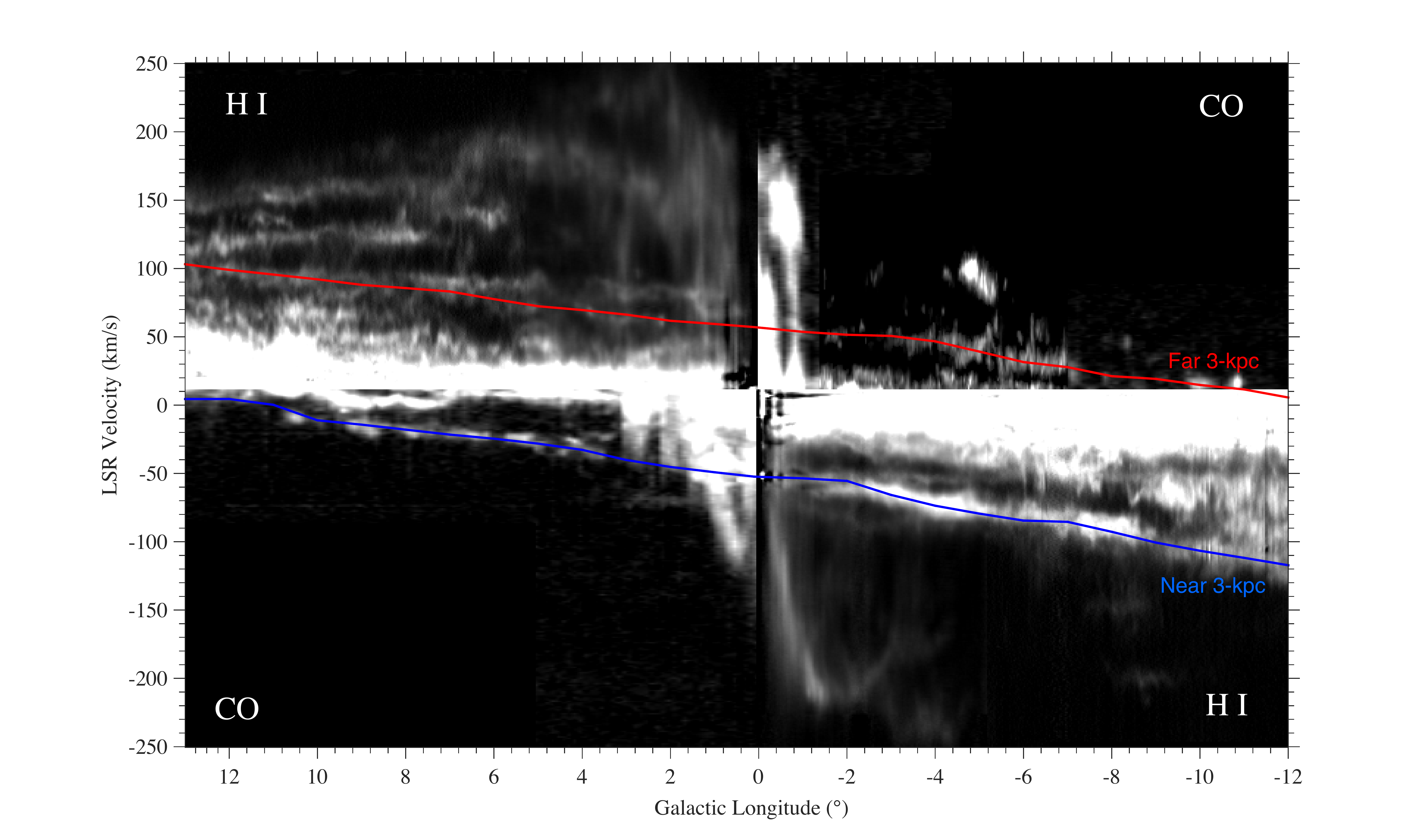}
\caption{\footnotesize
A composite of CO and \Hi\ surveys near $b=0\deg$ tracing the Near and Far 3-kpc arms.  
The Far 3-kpc arm is best traced in CO at negative longitudes, where it is closer to the far 
end of the bar and apparently richer in molecular gas; in \Hi\ this section of the arm is badly 
blended with \Hi\ emission from the foreground and outer disks. The mirror situation holds for 
the Near 3-kpc arm; it being best traced by CO at positive longitude and by \Hi\ at negative 
longitudes. The CO data are from the CfA survey, except in the range 
$l = -1.4\deg~{\rm to}~-7\deg$, where high-resolution observations with the Mopra telescope 
are used (Dame \& Barnes, private communication). 
The \Hi\ is from the SGP 21-cm survey of \citet{McClure-Griffiths:12}. The CfA survey CO 
emission is at $b = 0\deg$, while the higher resolution Mopra and SGP survey data are averaged 
over a strip of latitude roughly equal to the CfA beam of 8 arcmin.
        }
\label{fig:3kpc}
\end{figure}

\begin{figure}[ht]
\epsscale{0.85} 
\plotone{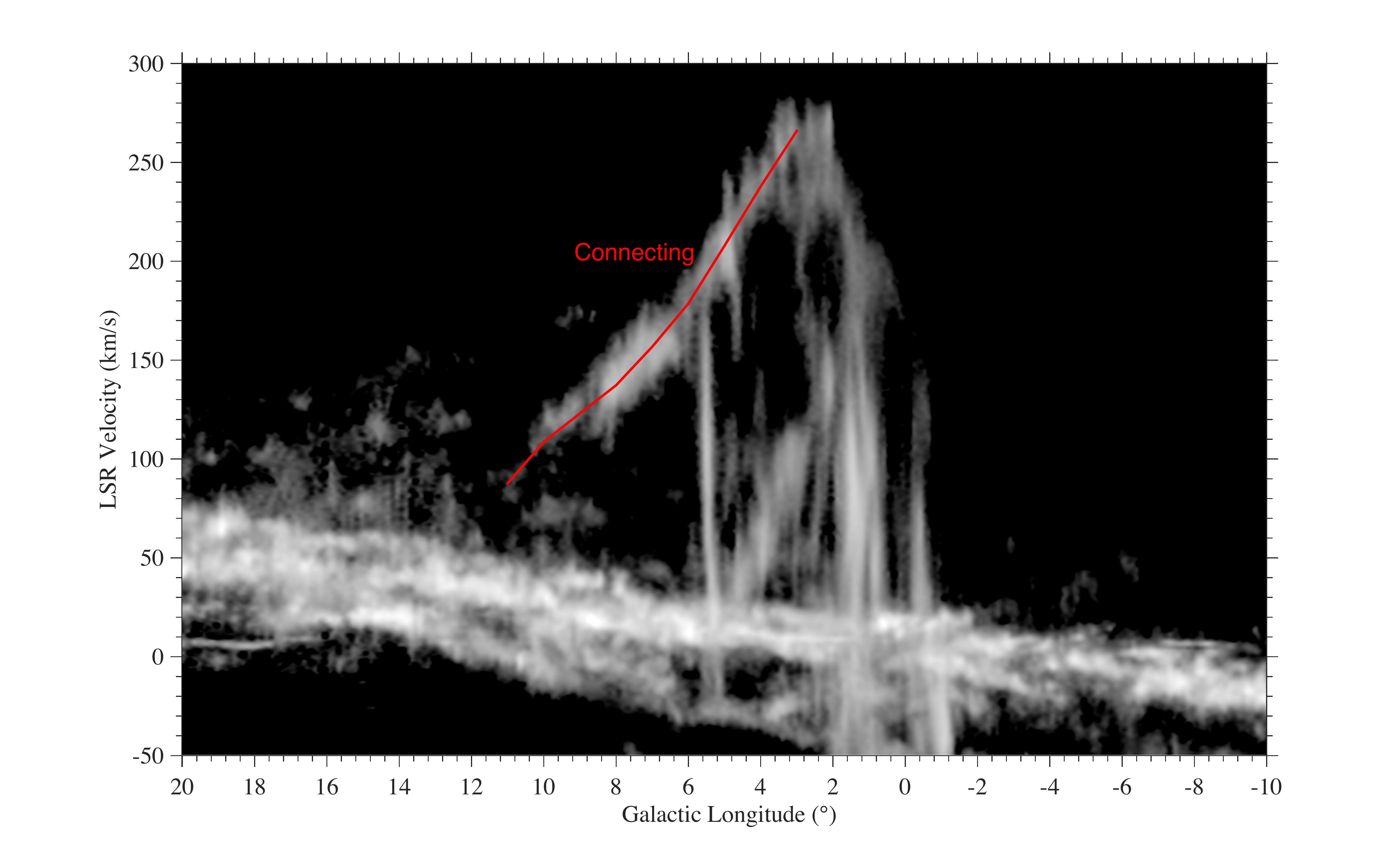}
\caption{\footnotesize
A trace of the Connecting arm on the CfA CO survey integrated from 
$b = -1.25\deg~{\rm to}~-0.25\deg$.
        }
\label{fig:Connecting}
\end{figure}

\begin{figure}[ht]
\epsscale{0.85} 
\plotone{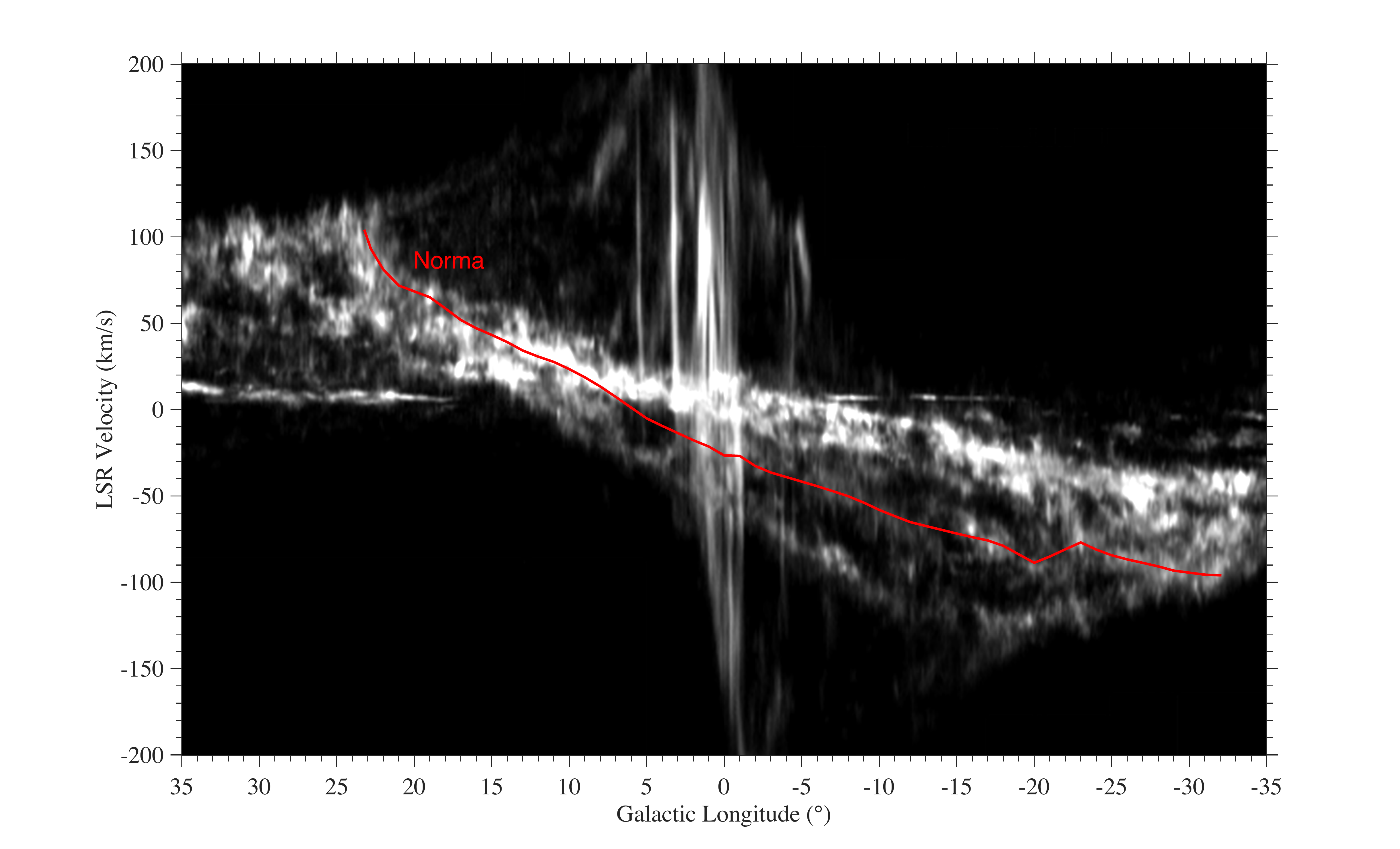}
\caption{\footnotesize
A trace of the Norma arm on the CfA CO survey integrated from $b = -1\deg~{\rm to}~+1\deg$.
        }
\label{fig:Norma}
\end{figure}

In Figs. \ref{fig:quad1}--\ref{fig:Norma} we optimally display and trace the numerous 
arcs and loops in HI 21 cm and CO \lv\ diagrams that have long been recognized as 
Galactic spiral arms \citep{Weaver:70,Cohen:80}. These features were used by the BeSSeL 
Survey to assign high mass star forming regions with measured parallaxes to spiral 
arms, and they can likewise be used for any 
spiral arm tracer with a measured velocity by our program.  While some of the 
features are very obvious (\eg  the Perseus arm in the second quadrant; 
Fig.~\ref{fig:q1q2_Perseus}), others are less so owing to blending with unrelated 
emission either at the opposite kinematic distance in the inner Galaxy or at other 
latitudes. For instance, the Far 3-kpc Arm \citep{Dame:08} can only be seen clearly 
in an \lv\ diagram integrated over a very narrow range of latitude near the plane 
(see Fig. \ref{fig:3kpc}). Other features are not only narrow in latitude but are 
offset from the plane and may even vary in latitude as a function of both longitude 
and velocity (\eg  the Outer Scutum-Centaurus arm; Fig. \ref{fig:q1q2_OSC}).  
Details on how the \lv\ diagrams were produced to best display each arm are given 
in the captions.

All of the spiral arm tracks, indicated by the colored lines in the figures, roughly 
correspond in shape and alignment with those expected for segments of logarithmic spiral 
arms, but we do not fit the features in that way, nor do we extrapolate or link up 
tracks in regions where they are not clearly seen. Rather, we run the tracks through 
the actual emission features that define the arms in longitude, latitude, and velocity.  
In some cases this results in a jaggedness that is slightly subjective, but they 
are the tracks of the actual spiral arms in the raw spectral-line data, and they are 
the structures that are now being located accurately in the Galaxy for the first time 
using maser parallaxes.  How or whether these structures link up into a grand design 
spiral pattern is not relevant for the present purpose.

\end{document}